\documentclass[%
 preprint,
 amsmath,amssymb,
 aps, 
prfluids,
]{revtex4-2}

\usepackage{graphicx}
\usepackage{dcolumn}
\usepackage{bm}
\renewcommand{\selectlanguage}[1]{}
\usepackage[mathlines]{lineno}

\begin{document}

\preprint{APS/123-QED}
\title{Development and application of a multiphase Lagrangian structure function model in anisotropic turbulence}

\author{Andrew P. Grace}  \email{Contact author: andrew.grace@colorado.edu}
\affiliation{Department of Applied Mathematics, University of Colorado Boulder, Boulder, Colorado 80309, U.S.A.\\}%

\author{David Richter}%
\affiliation{%
Department of Civil and Environmental Engineering and Earth Sciences, University of Notre Dame,
Notre Dame, Indiana 46556, U.S.A.\\
}%


\date{\today}

\begin{abstract}

The energetic response of inertial particles to turbulent flow motions is important for both a fundamental understanding of the multi-phase dynamics at play, and for applications such as reduced-order models of particle dispersion processes, and their two-way coupled effects onto the flow phase. Numerous studies focus on the energetics of ensembles of particles in homogeneous isotropic turbulence, where the influence of flow anisotropy (such as that provided by boundary conditions, or other external forcing) is not considered \textit{a priori}. In this work, we investigate the role of flow anisotropy on the Eulerian scale-wise particle phase energetics in a turbulent wall bounded flow for settling inertial Lagrangian particles. By using coupled Eulerian-Lagrangian direct numerical simulations at moderate Reynolds number, we aim to unravel the complex dependency of the scale-wise particle energetics on the turbulence intensity, particle inertia, and particle settling. In particular, we focus on how the developing anisotropy of the underlying turbulent flow (derived from the presence of the wall) is donated to the particle phase, and how particle inertia and settling preserve this large scale anisotropy into the formally isotropic scale range of the flow. We derive an exact (but unclosed) conservation law for the particle phase energetics at arbitrary scale, and use an asymptotic argument to help elucidate our DNS data. We discuss the relative changes to the quasi-streamwise and vertical components of the fluctuating particle field, and finish by discussing the implications of anisotropic non-local effects for more general flows, and the implications for continuum models of inertial settling Lagrangian particles.

\end{abstract}

\maketitle


\section{\label{sec:intro}Introduction}


Particle-laden turbulent flows are ubiquitous in the natural and engineered environment, and understanding the characteristics of these systems is important for many applications in the natural sciences and industry \cite{balachandar_turbulent_2010,brandt_particle-laden_2022}. Depending on the process in question, particles of interest can range in size from nanometer scales, to thousands of microns. As particles become larger, so too does their inertia, and it is known that inertial particle trajectories can differ significantly from the trajectories of fluid parcels in non-trivial ways \cite{bragg_relationship_2015}. One important implication of particle inertia is that particles may carry some memory of the flow motions that they have experienced in their history, leading to a dynamically interesting path-history effect. This path-history effect is implicitly linked to particle clustering and dispersion, which has many important implications for the transport of aerosols such as dust and droplets \cite{richter_inertial_2018,berk_dynamics_2021,berk_dynamics_2023}, their collision tendencies, and how energy is transferred spatially and spectrally within the solid phase. Properly understanding these small scale dynamical features of the solid phase is essential for accurately modeling the inherently multi-scale nature of their dynamics in a computationally efficient continuum model. 

Despite many existing studies focusing on energetics of particles in homogeneous isotropic turbulence (HIT), turbulent flows of practical interest often contain spatial inhomogeneities and anisotropies (directional dependence). The HIT model is possibly the simplest model for real turbulence, and allows investigators to probe the effects turbulent flow motions on particle dynamics in the dissipative range \cite{bec_heavy_2007}, and the inertial sub-range \cite{bragg_mechanisms_2015} in a controlled and isolated manner. However, inhomogeneity and anisotropy lend some important tendencies to particle dynamics that must be considered. In a wall-bounded turbulent flow such as a turbulent boundary layer (TBL), energy is extracted from a mean shear to feed turbulent fluctuations, which are damped by viscosity near the solid wall. Depending on the distance from the wall, these motions appear as large streak-like structures, known as large scale motions (LSMs), and at large enough Reynolds numbers, they are referred to as very large scale motions (VLSMs) \cite{smits_highreynolds_2011}. Thus, a solid boundary creates the conditions necessary for a (relatively) simple anisotropic and inhomogeneous turbulent flow; an increase in complexity compared to HIT. Importantly, as inertial particles carry the memory of larger scale fluctuations into smaller scale ranges, it is reasonable to suppose that if those larger scales are anisotropic, that they may carry some memory of the large scale flow anisotropy into the formally isotropic small scales. Therefore, for a formally isotropic small scale flow structure, the particle phase may be anisotropic due to the memory effect. 

An additional fundamental consideration is that near the solid boundary, turbulent fluctuations become weak, and one may not generally make the assumption that the flow fluctuations will possess accelerations large enough to overcome the accelerations due to a mean drift, such as gravitational accelerations. If we are considering a case with net particle deposition (i.e. the particles are, on average, traveling in the direction of gravity) the effect of the gravitational field cannot necessarily be disentangled from particle inertia \cite{brandt_particle-laden_2022}, since its effect is not simply additive, but is instead implicit to the particle dynamics. For example, recent work on settling inertial particle dynamics in turbulent boundary layers have revealed a complex coupling between gravity and inertia as particles enter the buffer layer and viscous sublayer of the wall bounded flow \cite{bragg_mechanisms_2021,grace_reinterpretation_2024}. As such, particle inertia and gravity must be considered simultaneously \cite{brandt_particle-laden_2022}. 

In this work, we investigate the scale-wise energetics of the particle phase in a turbulent boundary layer (TBL) under net deposition conditions using what are known as second order structure functions (abbreviated as SoSF), which quantify the energy contained below a longitudinal scale separation $r$. Our aim is to provide crucial insight into the scale-wise energy fluxes, and how the large scale anisotropy imparted on the solid phase by the fluid phase is preserved in the particle phase. In section \ref{sect:methods}, we introduce our measure of scale-wise flow anisotropy, $\eta^{(p)}$, in terms of the particle phase SoSF, and we use a kinetic theory approach to derive a continuum equation for this quantity. In section \ref{sect:results}, we compare $\eta^{(p)}$ and the particle phase SoSF to the fluid phase counterpart, and investigate their behavior numerically by performing Eulerian-Lagrangian direction numerical simulations (DNS) of a TBL (discussed in section \ref{sect:model}), and use an asymptotic argument to analyze the observed balance. Finally, we close with section \ref{sect:discussion}, which contains a summary and discussion highlighting the implications for continuum modeling of inertial settling particles, and how the preserved anisotropy may affect two-way coupled dynamics. 


\section{Energetic Characterization and Anisotropy Analysis Methods \label{sect:methods}}

To quantify the scale-wise energy distribution in both the solid phase and the fluid phase, and thus the particle phase anisotropy, $\eta^{(p)}$, we consider the second order longitudinal and vertical structure functions, which capture the energy (or more strictly speaking, the variance) contained in motion below a scale $r$. To facilitate a direct comparison between fluid and solid phases, we adopt an Eulerian view of the particle phase structure functions. For the solid phase, we consider the energy contained below a longitudinal scale $r$ in a continuum representation of the Lagrangian particle field. This will allow us to make a direct comparison to the fluid phase second order structure functions in sections \ref{sect:carrier} and \ref{sect:results}, which are an Eulerian quantity \textit{a priori}.

For the solid phase, we use a kinetic theory approach to derive the governing continuum equations for the second order structure functions. As these continuum equations will take into account the particle equations of motion (discussed below), they will allow us to understand the physical mechanisms that govern the energetic balance for a scale $r$. This analysis is outlined in section \ref{sect:particle}. Comparatively, to compare to the energetics for the fluid phase, we consider the scaling laws developed for second order structure functions outlined in \cite{davidson_logarithmic_2006}, which highlight how these quantities scale in the logarithmic region of a turbulent boundary layer. 

For both phases, we also compute these quantities using high resolution direction numerical simulations of one way coupled particle-laden wall bounded turbulent flow. The model setup can be found in section \ref{sect:model}.

\subsection{Particle Phase Energetic Characterization \label{sect:particle}}

In the following analysis, particles are implemented in a Lagrangian frame of reference. Solid-phase anisotropies have been studied in a TBL by considering the Lagrangian structure functions \cite{pitton_anisotropy_2012}, where the dependent variable is a time lag, instead of a spatial separation. Here, we make the comparison to the Eulerian fluid phase by instead considering the particle response from an Eulerian perspective. In the following section, we define the particle field via an approach leveraging kinetic theory \cite{reeks_development_2021}. First, we consider the conservation of momentum for small spherical particles immersed in a viscous fluid in the limit of a large particle-to-fluid density ratio. These are known as the Maxey-Riley equations, and are written
\begin{align}
    \frac{d}{dt}\boldsymbol{x}_p^{(r,s)} &= \boldsymbol{v}_p^{(r,s)}, \\
    \mathrm{St}^+\frac{d}{dt}\boldsymbol{v}_p^{(r,s)} &= \boldsymbol{u}(\boldsymbol{x}_p^{(r,s)}(t),t) - \boldsymbol{v}_p^{(r,s)} - \mathrm{Sv}^+\hat{\boldsymbol{z}}, \label{particle eoms}
\end{align}
where $\boldsymbol{x}_p$ represents the vector location of the particle centroid, and $\boldsymbol{v}_p$ represents the particle's vector velocity, and $\boldsymbol{u}$ represents the fluid velocity at the particle's location (referred to as the sampled velocity). We assume that the particle Reynolds number is small, meaning that the viscous drag along the surface of such particles is linear. For finite particle Reynolds number, one may consider a non-linear drag correction \cite{clift_bubbles_2005}, however this correction renders the derivation of the continuum equations intractable \cite{bragg_mechanisms_2021}, and is ignored in the following analysis. The superscript $(r,s)$ indicates whether or not a particle is a reference particle or satellite particle, respectively, and will be necessary to describe the statistics of particle separation in the following analysis. Finally, these equations are characterized by two governing parameters, $\mathrm{St}^+$, and $\mathrm{Sv}^+$, defined as 
\begin{equation}
    \mathrm{St}^+ = \frac{\tau_p u_\tau^2}{\nu}, \quad \mathrm{Sv}^+ = \frac{v_g}{u_\tau}.
\end{equation}
$\mathrm{St}^+$ is known as the Stokes number, which is the ratio between the particle's natural relaxation time, $\tau_p$, and the friction timescale $\nu/u_\tau^2$ ($\nu$ is the flow viscosity and $u_\tau$ is the flow friction velocity), whereas $\mathrm{Sv}^+$ is known as the settling parameter, and is a ratio of the laminar settling velocity, $v_g = \tau_p g$ (the theoretical speed with which a particle a settles through a viscous fluid under the action of gravity, $g$), and the friction velocity scale. The $+$ superscript indicates that dimensional quantities are scaled by these viscous quantities.

\begin{figure}
    \centering
    \includegraphics[width=\textwidth]{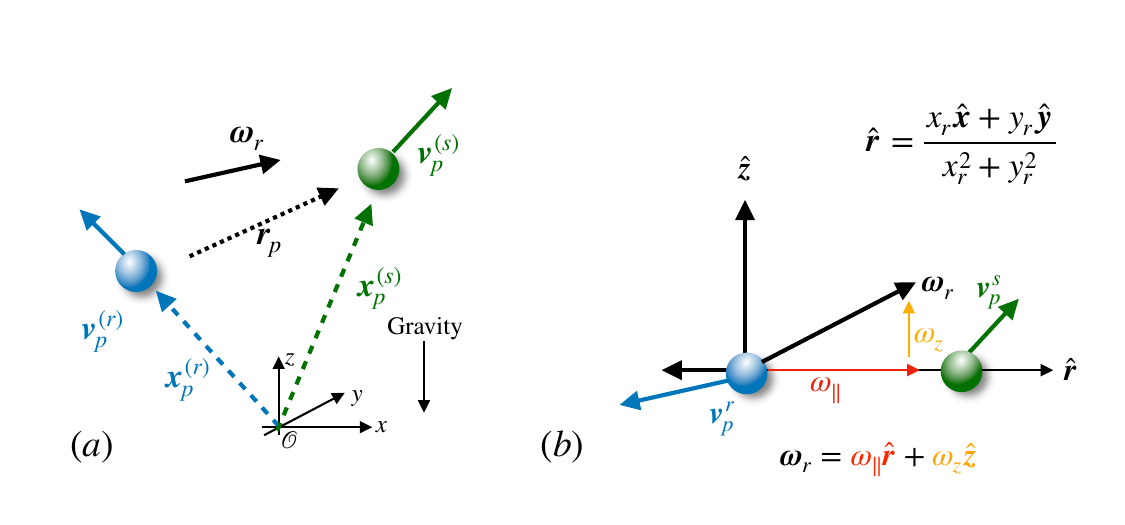}
    \caption{Panel (a) shows the location and velocity vectors of reference and satellite particles (superscript $(r)$ and $(s)$, respectively) in a fixed cartesian frame. $\boldsymbol{r}_p$ is the vector pointing from the reference particle to the satellite particle, $\boldsymbol{w}_r$ is the relative velocity vector of the two particles, and gravity points in the $-\hat{\boldsymbol{z}}$ direction. Panel (b) shows the reference and satellite particles in the frame of reference of the reference particle, and highlights the coordinate system used to represent the two-point quantities.}
    \label{fig:coords}
\end{figure}

Next, we define the time-dependent Lagrangian particle separation vector as $\boldsymbol{r}_p = \boldsymbol{x}^{(s)} - \boldsymbol{x}^{(r)}$, as well as the Lagrangian relative velocity vector of the particles, $\boldsymbol{\omega}=\boldsymbol{v}^{(s)} - \boldsymbol{v}^{(r)}$. Taking time derivatives of $\boldsymbol{r}_p$ and $\boldsymbol{\omega}$ and using \eqref{particle eoms}, the equations governing the evolution of the Lagrangian separation and relative velocities are
\begin{align}
        \frac{d}{dt}\boldsymbol{r}_p &= \boldsymbol{\omega}_p, \\
    \mathrm{St}^+\frac{d}{dt}\boldsymbol{\omega}_p &= \Delta\boldsymbol{u}(\boldsymbol{r}_p(t),\boldsymbol{x}_p^{(s)}(t),t) - \boldsymbol{\omega}_p.
\end{align}
We have illustrated this setup in figure \ref{fig:coords}(a).

Here, we have defined the relative sampled velocity as $\Delta \boldsymbol{u}(\boldsymbol{r}(t),\boldsymbol{x}^{(s)}(t),t)) = \boldsymbol{u}(\boldsymbol{x}^{(s)}(t),t)- \boldsymbol{u}(\boldsymbol{x}^{(r)}(t),t)$, and this is a function of the positions of the reference particle and the satellite particle, or equivalently, the position of the satellite particle and the separation between the reference and satellite particles. 
The equations governing the particle positions (velocities), i.e. equation \eqref{particle eoms}, share qualitative similarities with the relative positions (relative velocities), with one notable difference being the absence of the mean drift term for the particle relative velocities. However, this term implicitly affects the trajectories of the particles, and therefore the sampled relative velocity, $\Delta\boldsymbol{u}$. The separation vector and the relative velocity vectors are shown in the original frame of reference in figure \ref{fig:coords}(a) and in the frame of reference of the reference particle in figure \ref{fig:coords}(b). In this work, we shift the coordinate system from cartesian to cylindrical polar by making the coordinate change 
\begin{equation}
    \boldsymbol{\omega}_p = \omega_\|\hat{\boldsymbol{r}} + \omega_z\hat{\boldsymbol{z}}_r,
\end{equation}
allowing us to express the separation velocity in terms of longitudinal separation velocity $\omega_\|$, transverse separation velocity $\omega_\theta$, and vertical separation velocity $\omega_z$. 

We may consider these equations, along with \eqref{particle eoms}, as a dynamical system characterizing the two-particle evolution. To move to the Eulerian frame, we must consider an ensemble of these systems. We may define the joint probability distribution function (JPDF) $\mathcal{P} = \mathcal{P}(\boldsymbol{x},\boldsymbol{r},\boldsymbol{v},\boldsymbol{w})$ as 
\begin{equation}
    \mathcal{P} = \overline{\delta (\boldsymbol{x}_p^{(s)}  - \boldsymbol{x})\delta (\boldsymbol{r}_p  - \boldsymbol{r})\delta (\boldsymbol{v}_p^{(s)}  - \boldsymbol{v})\delta (\boldsymbol{\omega}_p - \boldsymbol{\omega})},
\end{equation}
where $\delta(\cdot)$ is the Dirac Delta function, and $\overline{\,\,\cdot\,\,}$ indicates an ensemble average over all turbulent realizations of the system. Here, $\boldsymbol{x}$, $\boldsymbol{v}$, $\boldsymbol{r}$, and $\boldsymbol{\omega}$ are the phase space variables. They are the spatial location and velocity of the satellite particles, the separation, and the relative velocity. Note that the JPDF can be equivalently defined in terms of the reference particle position and velocity in place of the relative position and relative velocity \cite{bragg_mechanisms_2015}. 

In this work, we consider ``slab-wise" statistics, where we condition our analysis on a zero vertical separation. Our practical approach for this conditioning is discussed more in section \ref{sect:model}. This simplification allows us to consider spatially local information in the vertical direction, i.e. we only consider satellite and reference particles located at the same height, meaning particle statistics are now conditioned only a height $z^{(r)}=z$. Overall, these simplifications reduce the functional dependence of the JPDF to $\mathcal{P} = \mathcal{P}(r,\theta,\boldsymbol{x},\boldsymbol{v},\boldsymbol{\omega})$ and importantly $\int_{\Omega} \mathcal{P}d\boldsymbol{\xi} = 1$ where $d\boldsymbol{\xi} = rdrd\theta d\boldsymbol{x} d\boldsymbol{v} d\boldsymbol{\omega}$ and $\Omega = [0,\infty)\cup [-\pi,\pi]\cup\mathbb{R}^3\cup\mathbb{R}^3\cup\mathbb{R}^3$ represents the domain of the phase space. 
Switching to index notation, the exact PDE governing the evolution of the JPDF is 
\begin{equation}
    \frac{\partial \mathcal{P}}{\partial t} + \frac{\partial}{\partial x_\alpha}v_\alpha\mathcal{P} + \frac{\partial}{\partial v_\alpha}\dot{v}_\alpha\mathcal{P}+ \frac{1}{r}\frac{\partial}{\partial r}r\omega_\|\mathcal{P} + \frac{\partial}{\partial \omega_\alpha}\dot{\omega}_\alpha\mathcal{P} = 0, \label{conservation equation}
\end{equation}
indicating that the JPDF is conserved in phase space.  
Note that in this work, Greek indices will imply the Einstein summation convection, whereas Latin indices do not.


Our interest lies in the second order structure functions, which are defined as 
\begin{equation}
   \langle {\omega'_i}^2\rangle =\frac{1}{\varrho}\int_{-\infty}^\infty \left(\omega_i - \langle\omega_i\rangle\right)^2\mathcal{P}\, d\boldsymbol{\omega}d\boldsymbol{v}, \label{second order structure func}
\end{equation}
where in the above, $i=\|,z$ is a short-hand for the longitudinal and vertical components, respectively. In this work, we do not consider transverse separations, and therefore do not consider the $\theta$ component. Transverse separation has been considered previously with regards to particle clustering in HIT and a TBL (see \citet{gualtieri_anisotropic_2009,wang_inertial_2019} for example).
The notation $\langle \cdot\rangle$ denotes an average conditioned on a longitudinal separation $r$ and height $z$. Note that in some work, the notation $\langle \cdot\rangle_{r,z}$ is used, where the subscripts explicitly indicate the quantities conditioned upon (see \cite{bragg_mechanisms_2021}, for example). We choose to simplify the notation in this work as the conditional average will always be upon $z$ and $r$ in this work, meaning the average is unambiguous. 

$\langle \omega_i\rangle$ represents the average relative velocities in each direction, which are not zero in general in this system, and $\varrho$ is defined as the marginal JPDF, i.e.
\begin{equation}
    \varrho(r,z) = \int_{\boldsymbol{w}}\mathcal{P}\, d\boldsymbol{\omega}d\boldsymbol{v},
\end{equation}
which represents the probability density that two particles are both located at height $z$, and are separated by a longitudinal distance $r$. This quantity is related to the horizontal radial distribution function \cite{bragg_new_2014}.

To compute conservation laws for $\langle {\omega'_i}^2\rangle$, we compute the second relative velocity moments of \eqref{conservation equation}. The computation is done explicitly in Appendix A. Here we re-produce the equations in compact form using index notation to highlight the structural dependence of the particle phase SoSF:
\begin{equation}
    \varrho\langle {\omega'_i}^2\rangle^+ = \varrho\langle \Delta u'_i\omega'_i\rangle^+ - \frac{1}{2}\mathrm{St}^+\left(\mathcal{D}_i  + \mathcal{M}_i + \mathcal{F}_i\right) \label{SoSF}. 
\end{equation}
The term $\langle \Delta u'_i\omega'_i\rangle^+$ represents the covariance between flow separation velocities and particle separation velocities. For $\mathrm{St}^+\rightarrow 0$, this term approaches the flow phase SoSF, but when particles possess a sufficiently large inertia, particle trajectories become de-coupled from flow streamlines, and their covariance approaches zero. There is also an explicit functional dependence on $\mathrm{St}^+$, and is composed of three parts. The term $\mathcal{D}_i$ is a linear transport operator that acts on the particle SoSF itself, meaning that the above equations are differential equations instead of algebraic equations, and is defined (in dimensionless terms) as 
\begin{equation}
        \mathcal{D}_i = \varrho\langle w\rangle^+ \frac{\partial}{\partial z^+}\langle {\omega'_i}^2\rangle^+ + \varrho\langle \omega_\|\rangle^+\frac{\partial}{\partial r^+}\langle {\omega'_i}^2\rangle^+,
\end{equation}
where $\langle w\rangle^+$ is a one point statistical quantity that represents the averaged vertical velocity of a set of particles a longitudinal distance $r$ from a reference particle (i.e. irrespective of the reference particle's velocity) and is formally defined as 
\begin{equation}
    \varrho \langle w\rangle = \int_{-\infty}^{\infty}w\mathcal{P}\, d\boldsymbol{v}d\boldsymbol{\omega}.
\end{equation}
$\mathcal{M}_{i}$ represents a coupling to the first order moments and appears as
\begin{equation}
    \mathcal{M}_i = 2\langle w'{\omega}'_z\rangle^+\frac{\partial}{\partial z^+}\langle \omega'_i\rangle^+ + 2\langle \omega'_i\omega'_\|\rangle^+\frac{\partial}{\partial r^+}\langle \omega_i\rangle^+,
\end{equation}
whereas $\mathcal{F}_i$ represents the particle phase energy flux and is expressed as
\begin{equation}
\mathcal{F}_i = \frac{\partial}{\partial z^+}\varrho \langle w'{\omega'_i}^2\rangle^+ + \frac{1}{r^+}\frac{\partial}{\partial r^+}r^+\varrho\langle \omega'_\|{\omega'_i}^2\rangle^+.
\end{equation}
Therefore, $\mathcal{F}_\|$ represents a coupling to higher order moments and renders \eqref{SoSF} unclosed. Importantly, $\mathcal{F}_i$ contains $\varrho$ within the derivative. The implication here is that if we were to take a derivative, $\mathcal{F}_i$ would contain a term proportional to the $r$ and $z$ derivatives of $\varrho$, and one to $\varrho$ itself. This kind of term is often interpreted as having a diffusion tendency and a drift tendency \cite{bragg_new_2014-1}. 

Importantly, since the terms comprising the conservation equations for $\langle{\omega'_\|}^2\rangle$ are not equal to those comprising $\langle{\omega'_z}^2\rangle$ in general, an anisotropy in the particle phase is created. We define the metric quantifying the anisotropy for the particle phase as:
\begin{equation}
        \eta^{(p)}(r,z) =  \frac{\langle {\omega'_\|}^2\rangle^+ - \langle {\omega'_z}^2\rangle^+}{\langle {\omega'_\|}^2\rangle^+ + \langle {\omega'_z}^2\rangle^+}. \label{particle_aniso}
\end{equation}
When $\eta^{(p)}\rightarrow 1$, particle phase ``eddies" are pancake-like structures, with horizontal fluctuations being much stronger than vertical ones, while converse is true for $\eta^{(p)}\rightarrow -1$. Furthermore, when $\eta^{(p)}=0$, horizontal and vertical fluctuation magnitudes are equal and the particle phase is isotropic. Finally, the particle phase SoSF are defined such that $r\rightarrow \infty$, $\eta^{(p)} \rightarrow \eta^{(p)}_\infty$ where $\eta^{(p)}_\infty$ is the anisotropy of the integral scale.


\subsection{Flow Phase Energetic Characterization \label{sect:carrier}}
In this section, we review the theory underscoring the flow energetics TBL, specifically the scaling behavior of the flow phase SoSF, and use these to define the flow phase anisotropy. The goal here is to provide the basis through which we will discuss the particle field anisotropy in section \ref{sect:particle}. Turbulent boundary layers exhibit horizontal homogeneity, leaving mean quantities to be dependent only on height, $z$. As such, we can define $x_h = (x^2 + y^2)^{1/2}$ as the position in the horizontal plane, and $u_h = (u^2 + v^2)^{1/2}$ as the magnitude of horizontal fluctuations. As both $z$ and $\mathrm{Re}_\tau$ change independently, $u_h\rightarrow u$ since it is thought that outer scale motions tend to couple with inner scale motions in the horizontal whereas the spanwise and vertical components saturate \cite{kunkel_study_2006,pope_turbulent_2015,smits_highreynolds_2011}.

To quantify the energetics of the flow in this work, we rely on the horizontal two-point statistics at a fixed height $z$. That is, we quantify the the variance (and therefore the energy) contained in each component of the fluctuating velocity field up to a horizontal separation $r$. The SoSFs are defined for the horizontal and vertical components as
\begin{equation}
    \langle \delta u_h^2\rangle = \langle (u_h(x_h + r) - u_h(x_h))^2\rangle, \quad  \langle \delta w^2\rangle = \langle (w(x_h + r) - w(x_h))^2\rangle,
\end{equation}
which quantify the variance of the flow field in the horizontal and vertical directions below a scale horizontal scale $r$, and therefore the kinetic energy of flow motions in scales below $r$. Note that the interpretation of the second order structure function is different than that of the kinetic energy spectra (an alternative way of measuring the scale-wise energy), which gives a measure of the kinetic energy containing between wavenumbers $k$ and $k+dk$. Thus, the second order structure function is a strictly non-decreasing function which asymptotes to the total energy contained in all horizontal scales. 

\subsection{Scaling of the fluid phase second order structure functions}
Our primary focus in this work is analyzing settling particle dynamics within the logarithmic layer and the buffer layer. Below, we qualitatively analyze the dynamics in these two regions.
The scaling behavior of the horizontal components of the SoSF in the logarithmic region of a turbulent boundary layer are documented in \cite{davidson_logarithmic_2006,davidson_simple_2009,davidson_refined_2006,davidson_universal_2014}, whereas the horizontal structure of the vertical component is often not considered. Instead of an exhaustive analysis, the reader is referred to the aforementioned work by Davidson and collaborators for a detailed description, and here, we only touch on the salient details for the purpose of this work. The SoSF in the logarithmic region of the TBL takes on different scaling behavior depending on the horizontal scale of the fluctuations. The relevant length scales are the Kolmogorov microscale $\eta_k$, the local eddy scale $z$, and the boundary layer outer scale proportional to its total height $h$. These three length scales correspond to four scaling regimes: the dissipative range ($r\ll \eta_k)$, the inertial subrange ($\eta_k\ll r\ll z$), the logarithmic range, ($z<r\ll h$), and the uncorrelated range ($r>h$). 

At large enough Reynolds number and far enough from the influence of the solid boundary, we expect both the dissipative and inertial range dynamics to be identical for the horizontal and vertical components of the SoSF, as those motions have a short enough overturning times that they ``forget" the influence of the wall (i.e., they behave isotropically). Upon normalizing $r$ by the local large scale eddy size $z$ and expressing the structure function in terms of wall units, for the logarithmic region of the flow, it can be shown that in the dissipative scaling regime appears as
\begin{equation}
    \langle \delta u^2\rangle^+ \sim  \langle \delta w^2\rangle^+ \sim \left(\frac{z^+}{15\kappa}\right) \left(\frac{r}{z}\right)^{2}  \quad\frac{r}{z} \ll\kappa^{1/4}{z^+}^{-3/4}, \label{flow diss range}
\end{equation}
where $\kappa$ is the von K\'{a}rm\'{a}n constant.
Likewise, in the inertial sub-range (should it exist), it can be shown that 
\begin{equation}
\langle \delta u^2\rangle^+ \sim  \langle \delta w^2\rangle^+ \sim \left(\frac{C_0^{3/2}}{\kappa}\right)^{2/3} \left(\frac{r}{z}\right)^{2/3} \quad \kappa^{1/4}{z^+}^{-3/4} \ll \frac{r}{z} \ll 1,
\end{equation}
where $C_0$ is an $\mathcal{O}(1)$ constant. These results are highlighted in \cite{davidson_logarithmic_2006}.

Once in the logarithmic scaling region $z<r\ll h$ (not to be confused with the logarithmic region of the flow, $\nu/u_\tau \ll z \ll h$), the flow motions are no longer isotropic. Davidson and colleagues argue that a logarithmic scaling range occurs for the horizontal component. This is a consequence of the fact that horizontal motions in the overlap region between $z$ and $h$ are influenced by both their distance from the boundary as well as the outer scale of the boundary layer. It can be shown by using boundary layer matching techniques and field observation that the kinetic energy spectra in this overlap region must scale as $k^{-1}$ \cite{Perry_theoretical_1986,nickels_evidence_2005}, and by relating the spectra to the SoSF, the result is a logarithmic scaling law \cite{davidson_logarithmic_2006}. 

As far as we can tell, \citet{davidson_logarithmic_2006} do not specifically mention the scaling behavior of $\langle \delta w^2\rangle^+$ in their work. However, the vertical component becomes independent of the outer scale within the logarithmic layer at asymptotically large Reynolds number \cite{kunkel_study_2006}, thus implying no overlap with the outer scale, and no logarithmic scaling range in the vertical SoSF. Therefore, we match to the integral scale behavior for $r>z$, which is thought to be a constant for asymptotically high Reynolds number. 
Specifically, we take
\begin{equation}
    \langle \delta u^2\rangle^+ \sim A_1 + B_1\log\left(\frac{r}{z}\right),\,\, \langle \delta w^2\rangle^+ \sim A_0,\quad \quad 1<\frac{r}{z}\ll\frac{h}{z}, \label{flow log range}
\end{equation}
where $A_0$, $A_1$, and $B_1$ are $\mathcal{O}(1)$ constants, specified in \cite{davidson_simple_2009}. 

Finally, for $r>h$, we expect motion in both components to be completely uncorrelated, which brings us to the expressions for $2\langle \delta u^2\rangle^+ \sim \langle {u'}^2\rangle$ and $2\langle \delta w^2\rangle^+ \sim \langle {w'}^2\rangle$ so that, at asymptotically high Reynolds number,
\begin{equation}
    \langle \delta u^2\rangle^+ \sim A_2 + B_2\log\left(\frac{h}{z}\right),\,\, \langle \delta w^2\rangle^+ \sim A_0 \quad \frac{h}{z}<\frac{r}{z}, \label{integral flow dynamics}
\end{equation}
where $A_2$, and $B_2$ are $\mathcal{O}(1)$ constants defined in \cite{kunkel_study_2006}.

Strictly speaking, the above analysis is relevant for the logarithmic layer of the flow only. In regions of the logarithmic layer near the buffer layer, the size of the inertial sub-range may be small or even non-existent, meaning that some of the large scale anisotropy in the logarithmic range may ``leak" into the dissipative range, resulting in dissipative range anisotropy at certain heights, especially in the buffer layer where characteristic flow quantities scale with neither $\nu/u_\tau$ nor $z$. For the buffer region, we do not attempt to derive analytical scaling laws, and instead appeal to DNS.

We quantify the flow phase scale anisotropy in an identical way to \eqref{particle_aniso}, denoted by $\eta^{(f)}(r,z)$, defined as:
\begin{equation}
    \eta^{(f)}(r,z) =  \frac{\langle\delta u_h^2\rangle^+ - \langle \delta w^2\rangle^+}{\langle\delta u_h^2\rangle^+ + \langle \delta w^2\rangle^+}. \label{flow anisotropy}
\end{equation}
Within the logarithmic layer, the theoretical scaling behavior of the fluid phase SoSFs imply that the ansiotropy should go to zero within the dissipative and inertial sub-range scaling regimes, whereas at larger separations (specifically when $r>z$), the anisotropy should grow until is saturates at $r\gg h$.

\subsection{Numerical Model Setup\label{sect:model}}
\begin{figure}[h]
    \centering
    \includegraphics[width=\textwidth]{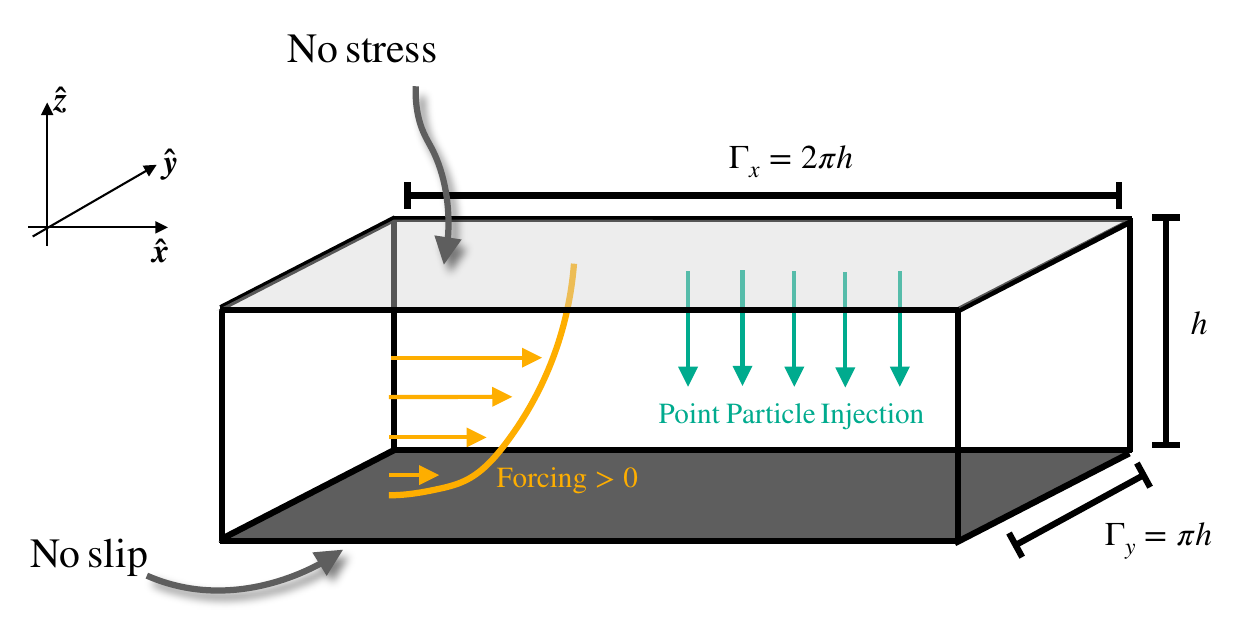}
    \caption{A schematic of the simulation domain. A pressure gradient is applied to generate the mean flow, which creates turbulence due to the friction at the no-slip lower boundary. Point particles are injected with at the no-stress upper boundary and are allowed to come to statistical equilibrium with the flow. Particles exiting the domain through the lower boundary are injected in a random horizontal location at the upper boundary to maintain a constant particle number, while particle reflect off the upper wall. Particles exiting through the periodic side boundaries are placed at the opposite wall while maintaining their velocity.}
    \label{fig:schematic}
\end{figure}
\subsubsection{Flow Phase Model}
We use the NCAR Turbulence with Lagrangian Particles Model \citep{richter_inertial_2018} to perform DNS of the incompressible Navier-Stokes equations one-way coupled to inertial Lagrangian particles in a turbulent open channel flow. The equations solved are 
\begin{align}
    \frac{\partial \boldsymbol{u}}{\partial t} + \boldsymbol{u}\cdot \nabla\boldsymbol{u} &= -\nabla p +\frac{1}{\mathrm{Re}_\tau}\nabla^2\boldsymbol{u} + \hat{\boldsymbol{x}} \\ 
    \nabla \cdot \boldsymbol{u} &= 0,
\end{align}
where $\mathrm{Re}_\tau$ is the friction Reynolds number, $\boldsymbol{u} = (u,v,w)$ is the three dimensional velocity vector evaluated at the cartesian coordinates $\boldsymbol{x} = (x,y,z)$ (streamwise, spanwise, vertical), and $\hat{\boldsymbol{x}}$ represents a unit vector in the streamwise direction. The forcing (of unit magnitude) is required to achieve a statistically steady flow state. The numerical setup is schematized in \ref{fig:schematic}.
The domain is horizontally periodic, and the flow is subjected to a no-slip boundary condition along $z = 0$, and a no-stress boundary condition along $z=h$ (i.e. the maximum height of the domain). Since we are focused on the dynamics of the logarithmic and buffer regions, the upper boundary condition does not affect the dynamics in a meaningful way. Due to computational restrictions, we focus on cases with a relatively low fixed $\mathrm{Re}_\tau=315$ so as to accommodate enough Lagrangian particles to achieve statistical convergence at the small scales. 

\begin{figure}
    \centering
    \includegraphics[width=\textwidth]{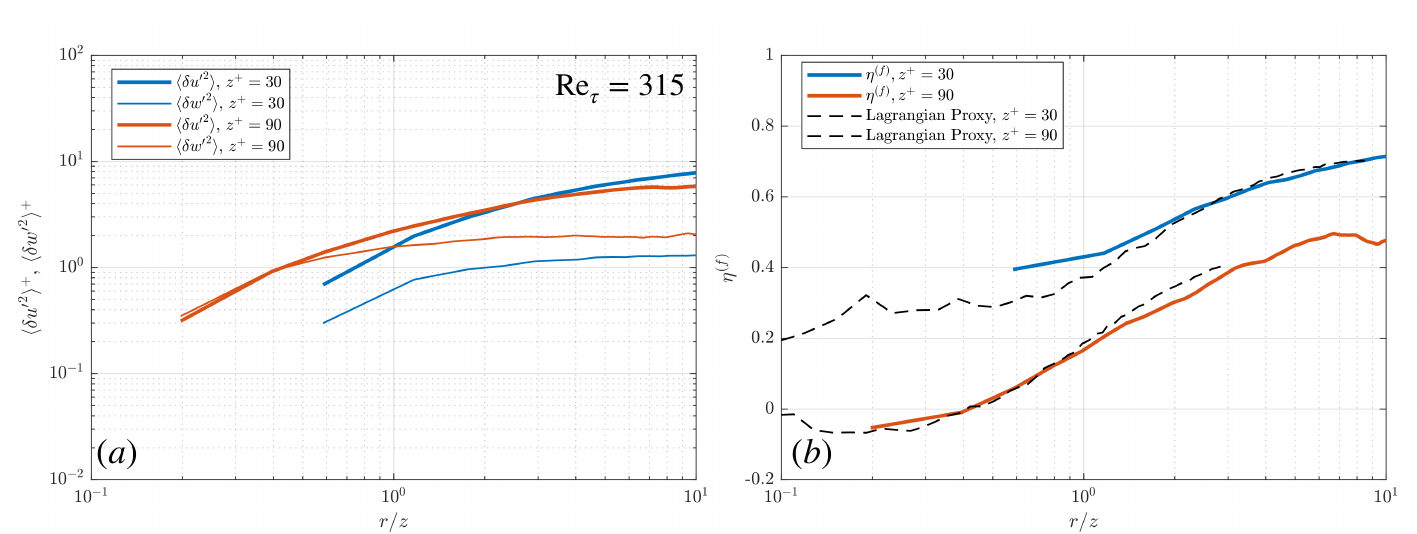}
    \caption{Panel (a) shows curves highlighting the horizontal and vertical components of the fluid phase SoSF at $z^+=30,\,90$ at fixed $\mathrm{Re}_\tau=315$. Panels (b) shows the flow phase anisotropy measure at the same heights (colored curves) overlaid with the Lagrangian proxy (discussed in section \ref{sect:particle model}). }
    \label{fig:SoSF_flow}
\end{figure}

As an example, Figure \ref{fig:SoSF_flow}(a) shows the normalized second order structure functions of the horizontal and vertical fluctuations at two different heights for $\mathrm{Re}_\tau = 315$ from DNS. These heights are representative of the logarithmic region of the flow, $z^+=90$, and the near-wall region, which is theoretically independent of the outer scale dynamics, $z^+=30$. For the logarithmic layer curves (orange), the nature of the curves above and below $r/z \approx 1$ is different, and is a reflection of the dissipative range and logarithmic range scaling laws respectively, (i.e. \eqref{flow diss range} and \eqref{flow log range} respectively). Specifically, in the dissipative range, the curves approximately collapse onto one another, suggesting good agreement with our hypothesis above that $\eta^{(f)}\rightarrow 0$ (shown in figure \ref{fig:SoSF_flow}(b)) in the limit of $r\ll z$, whereas they separate from one another for $r>z$, and will eventually saturate for $r\gg h$. 

Moreover, figure \ref{fig:SoSF_flow}(a) shows the fluid phase SoSFs, highlighting that there is no obvious inertial sub-range in at these heights. We note that at the heights we are concerned with, the size of the inertial subrange will be relatively small irrespective of the Reynolds number. The size of the inertial subrange scales as $z/\eta_k \sim \kappa^{-1/4}\left(z^+\right)^{3/4}$, so that the size of the inertial subrange is fixed for fixed $z^+$ and is independent of the outer scale. In fact, the lack of an inertial subrange is an important feature for this system and its impact will be discussed later in this work.

Finally, for the near-wall region, $z^+=30$ (blue curves), we see that the curves do not collapse on one another in the dissipative region, suggesting both small and large scale anisotropy as we approach the wall, i.e. $\eta^{(f)}>1$ as $z/h\ll 1$. This is expected as the scalings above are formally valid within the logarithmic layer of the flow, specifically where kinetic energy dissipation roughly balances shear production of turbulence \cite{pope_turbulent_2015}. Importantly, this balance is not generally the case at $z^+=30$. In Section \ref{sect:results}, we contrast the anisotropy of the fluctuating particle velocity field, and use the insight gained from the above analysis to investigate how the anisotropies donated to the fluctuating particle velocity field by the carrier phase are modified by the particle's inertia and mean drift due to gravity (i.e. $\mathrm{St}^+$ and $\mathrm{Sv}^+$).


\subsubsection{Particle Phase Model \label{sect:particle model}}

The particle treatment occurs via the Maxey-Riley equations for a small rigid spheres subjected to linear drag and the gravitational force, and are represented by equation \eqref{particle eoms} shown in Section \ref{sect:particle}. The particles are assumed to be non-interacting, small (relative to the local Kolmogorov scales), monodisperse, and have no effect on the flow state (i.e. one-way coupling). Therefore, the results contained within this work are relevant for dilute suspensions of dense but small spherical particles, such as in the atmospheric surface layer \cite{berk_dynamics_2021,berk_dynamics_2023}. Particles are removed when their centroid passes below $z=0$ and re-injected at a random horizontal location along $z=h$. This setup has been used frequently \cite{bragg_mechanisms_2021,grace_reinterpretation_2024,grace_effects_2025} in order to maintain both a total particle number in the domain, as well as a constant downward flux. 
We have also included the flow structure functions sampled by particles with $\mathrm{St}^+=0.1$ and $\mathrm{Sv}^+=0.025$ (dashed black curves) in figure \ref{fig:SoSF_flow} as our Lagrangian proxy for the flow structure functions as a comparison, and we refer to these values as ``Unconditional" in figure legends in section \ref{sect:results} below. These are included as they provide some information regarding the flow velocities sampled by the particles below the grid scale (due to the linear interpolation scheme), which the Eulerian flow velocities cannot show.

\section{Results \label{sect:results}}

\subsection{Results from Direct Numerical Simulation \label{sect:numerics}}
\begin{figure}
    \centering
    \includegraphics[width=\textwidth]{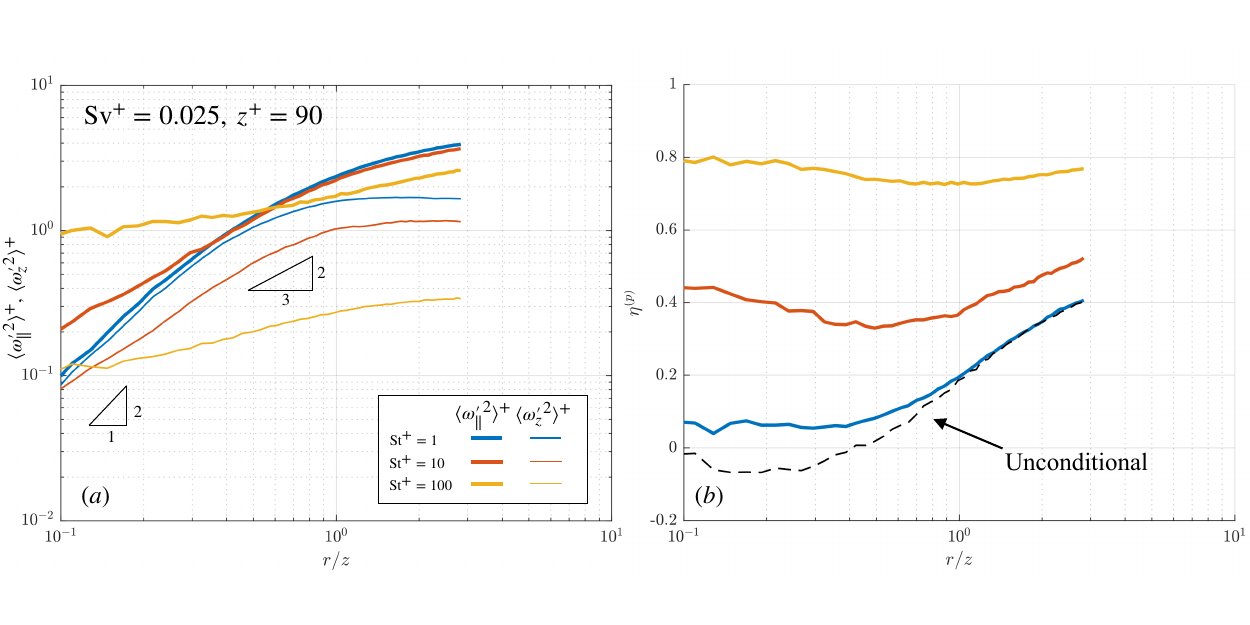}
    \caption{Panel (a) shows particle phase horizontal (thick solid curves) and vertical (thin solid curves) second order structure functions at $z^+ = 90$ at fixed $\mathrm{Sv}^+$.  Panel (b) shows particle phase anisotropy, $\eta^{(p)}$. Curve colors correspond to different values of $\mathrm{St}^+$.  Inertial range and dissipative range slopes are indicated on each figure.}
    \label{fig:StCompare}
\end{figure}

Our primary focus of this work is to identify the broad role of settling perturbations on the particle field energetics, and especially on how the large scale anisotropy may be reflected in the particle phase at smaller scales (i.e., smaller separations). However, before we move into an analysis of the modification of $\mathrm{Sv}^+$, it is useful to examine the overall qualitative characteristics of the particle field at fixed $\mathrm{Sv}^+$ and varying $\mathrm{St}^+$. Figure \ref{fig:StCompare}(a) shows the longitudinal and vertical particle structure functions for three different values of $\mathrm{St}^+$, while figure \ref{fig:StCompare}(b) shows the particle phase anisotropy. These quantities are plotted at a height of $z^+=90$ and against $r/z$ (i.e., the logarithmic scaling). We have also plotted slopes as a point of reference to qualitatively analyze the particle structure function behavior relative to expected flow phase scaling laws.

As particle inertia becomes large, the particle phase acquires much larger energy at small scales ($r/z<1$) relative to the flow. These non-local effects are evident in figure \ref{fig:StCompare}(a) but the magnitude of the increase is dependent upon the particle response to the horizontal or vertical fluctuations. The differences in this energetic content are also evident in the particle phase anisotropy, figure \ref{fig:StCompare}(b). At small Stokes numbers (i.e. $\mathrm{St}^+=1$) and small separations (approximately emulating the small scale fluid phase energetics), fluctuations are roughly of similar magnitude, suggesting isotropy, confirmed by the fact that $\eta^{(p)}$ approaches zero in figure \ref{fig:StCompare}(b) as $r/z\rightarrow 0$. Conversely, at larger Stokes numbers, there is a more significant difference between the longitudinal and vertical fluctuations across all scales, leading to a large degree of small-scale anisotropy in the particle phase, while the flow phase remains isotropic. Figure \ref{fig:StCompare}(b) also shows that the small scale anisotropy is at least as significant as the large scale anisotropy, even for for $\mathrm{St}^+=10$, suggesting that the memory effects of particle inertia preserve the large scale anisotropy in some way. 

\begin{figure}
    \centering
    \includegraphics[width=\textwidth]{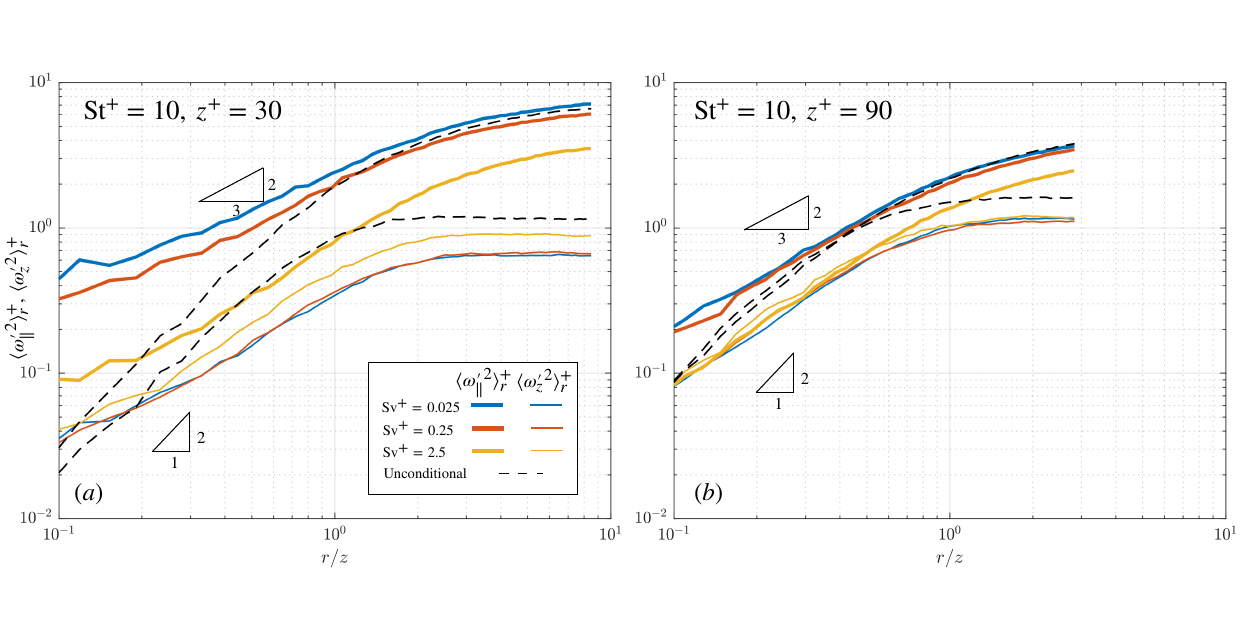}
    \caption{Particle phase horizontal (thick solid curves) and vertical (thin solid curves) second order structure functions at $z^+ = 30$ (a) and $z^+=90$ (b) for $\mathrm{St}^+=10$. Curve colors correspond to different values of $\mathrm{Sv}^+$. Unconditional (fluid phase) structure functions are plotted as dashed curves. Inertial range and dissipative range slopes are indicated on each figure.}
    \label{particle_structure_functions_at_heights}
\end{figure}

The role of particle inertia on the structure functions has been investigated by many studies in HIT \cite{bragg_new_2014-1, ireland_effect_2016}, and the qualitative findings of those studies follow for the axisymmetric particle dynamics in a wall bounded flow. Therefore, for the remainder of this work, we fix the particle Stokes number at $\mathrm{St}^+=10$, and vary $\mathrm{Sv}^+$ to foster insight into how settling, paired with moderate inertia, modifies the small scale particle phase anisotropy. Figure \ref{particle_structure_functions_at_heights} shows the horizontal and vertical structure functions at two different heights: $z^+ = 30$ and $z^+=90$. These two heights were chosen as representative examples of the buffer layer and the logarithmic layer flow dynamics. Note that we cannot choose $z^+$ much larger than 90 owing to the relatively low Reynolds number. For larger choices of $z^+$, there is likely to be some influence from the upper boundary condition (and thus the domain height) on the self-similar characteristics of the logarithmic region. However, we believe that $z^+=90$ is a representative example of the qualitative logarithmic layer dynamics expected at higher Reynolds numbers. 
 
Figure \ref{particle_structure_functions_at_heights} shows the modification of the particle structure functions relative to the flow state (i.e. the unconditional structure functions) as $\mathrm{Sv}^+$ is changed, and therefore provides some insight into how the particle phase energetics are modified by particle settling.
The particle phase structure function curves are strongly modified by particle settling within the buffer layer, shown in figure \ref{particle_structure_functions_at_heights}(a), with smaller changes appearing within the logarithmic layer, shown in figure \ref{particle_structure_functions_at_heights}(b). In common to both heights is the fact that the longitudinal components decrease minimally across all scales until $\mathrm{Sv}^+$ surpasses unity when there is a dramatic drop in magnitude. Interestingly, the vertical component actually increases back towards the unconditional value in the buffer layer whereas it remains fixed in the logarithmic layer. Moreover, at large $\mathrm{Sv}^+$, the structure functions collapse together within the logarithmic layer (indicating small scale isotropy for these cases) whereas they do not collapse in the buffer layer (indicating some preserved anisotropy at small scale). One potential explanation for this behavior is that the particle phase at large $\mathrm{Sv}^+$ tends towards the underlying flow anisotropy. As the small scales within the buffer layer are themselves anisotropic, the particle phase simply tends towards an anisotropic state. Conversely, the small scales within the logarithmic layer are isotropic (even at the relatively low Reynolds number here; see figure \ref{fig:SoSF_flow}(a)), so the state the particle phase tends towards at large $\mathrm{Sv}^+$ is an isotropic one.

The slopes on figure \ref{particle_structure_functions_at_heights} indicate that as the flow approaches a dissipative scaling range (i.e., $r^2$), the inertial particle structure functions exhibit a shallower inertial-range like scaling (i.e. $r^{2/3}$). This behavior is a reflection of the scale non-locality of particles and is driven primarily by particle inertia, and modulated by particle settling. For example, see the work of \citet{ireland_effect_2016-1} that highlights this scale non-locality in homogeneous isotropic turbulence for settling particles. Importantly, this basic qualitative result is preserved in both the buffer and logarithmic layers across all values of $\mathrm{Sv}^+$ considered. However, while both the longitudinal and vertical structure functions sharing a similar scaling, it is clear that both $\mathrm{Sv}^+$ and $z^+$ control their displacement from one another, revealing a complicated coupling between $\mathrm{St}^+$, $\mathrm{Sv}^+$, and $z^+$.  

\begin{figure}
    \centering
    \includegraphics[width=\textwidth]{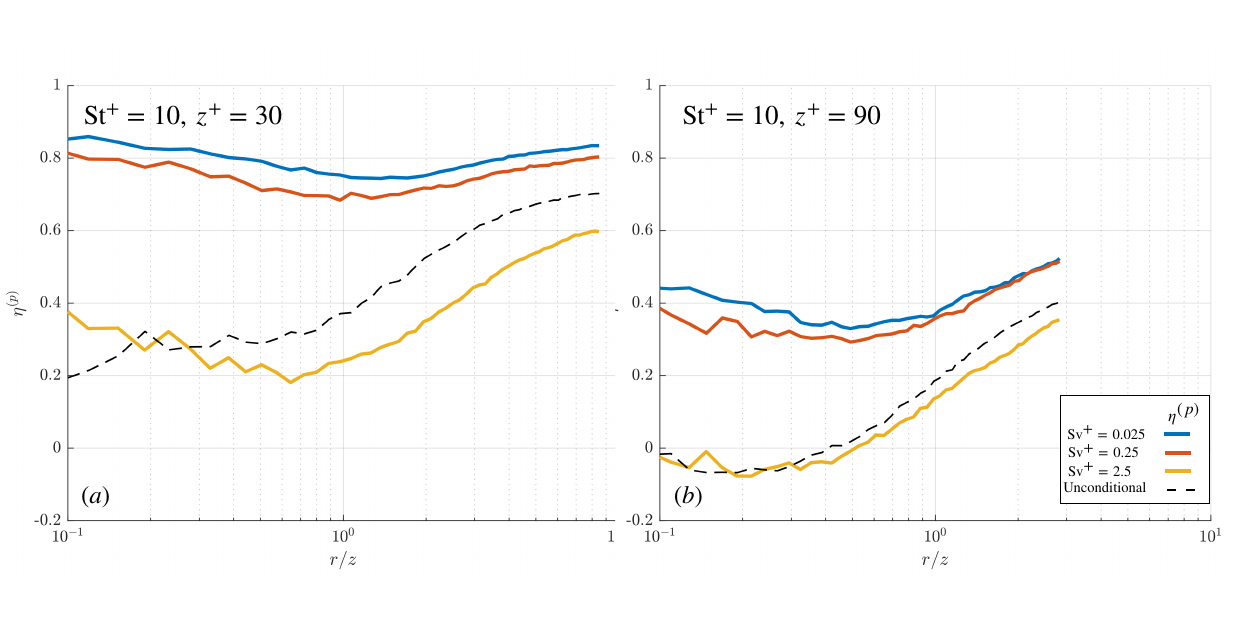}
    \caption{Particle phase anisotropy metric, $\eta^{(p)}$ at  $z^+ = 30$ (a) and $z^+=90$ (b) for $\mathrm{St}^+=10$. Curve colors correspond to different values of $\mathrm{Sv}^+$. Unconditional (fluid phase) anisotropy is plotted as dashed curves for each height.}
    \label{fig:eta_heightcompare}
\end{figure}

Figures \ref{fig:eta_heightcompare}(a) and (b) demonstrate the scale-wise anisotropy within the buffer layer and logarithmic layers, respectively, with the flow phase anisotropy indicated by the dashed curve (i.e. the Lagrangian proxy). Examination of these curves reveals a fundamental difference between the solid and fluid phases within both regions of the flow. For both regions under consideration, the flow phase anisotropy asymptotes towards some minimum value. In the buffer layer ($z^+ = 30$), the flow anisotropy  asymptotes towards a value around 0.2 as $r\rightarrow 0$, indicating that the buffer layer dynamics are not isotropic since the presence of the wall significantly affects the dynamics. Within the logarithmic layer ($z^+=90$), the flow recovers isotropy at small scales and approaches a value around 0 (though it is slightly negative here, likely due to the numerical scheme). 

Comparatively, the particle phase shows fundamentally different characteristics due to the scale non-locality. The particle phase for $\mathrm{Sv}=0.025$ and 0.25 hit a local minimum, and then begin to increase once again as $r/z \rightarrow 0$. This is broadly consistent across both values $z^+$, but the value of the minimum, and the scale at which it occurs are modified by these parameters. Interestingly, for $\mathrm{Sv}^+=2.5$, the large scale anisotropy is decreased, but the small scale anisotropy is reminiscent of the fluid phase, suggesting that for a large enough influence by gravity, the small scales may recover towards an isotropic state. However, as discussed above, the ``isotropic recovery" could be compensated in some way by the fact that the flow is itself somewhat anisotropic, especially in the buffer layer.

The non-monotonicity of these curves suggests the competition between (at least) two different mechanisms. At the large scales, the particle phase anisotropy is governed (at least in part) by the underlying anisotropy of the flow itself, meaning that as the flow phase anisotropy increases, so too does the particle phase anisotropy, suggesting some Reynolds number dependence. At the small scales, the story appears more complicated, since particles become increasingly insensitive to flow fluctuations, so the anisotropy must be governed by the inter-scale energy transfer of particle phase; a persistent energy flux from the large scales to the small scales. In the following section, we use the continuum theory derived in section \ref{sect:particle} as a lens through which we analyze these results. 

\subsection{Interpretation via asymptotic analysis of the continuum equations}

The goal of the following analysis is not to provide a predictive theory, but instead to provide deeper insight into the DNS results and to understand the underlying dynamics of the small scales of the particle phase. An approximate theory is required here due to the unclosed nature of the continuum equations \eqref{SoSF} (i.e. due to $\mathcal{F}_i$, which contains the third order moments of the velocity increments). 

We make the following power law assumptions for the mean two point quantities in the range $r/z\ll 1$. Note that we leave the lower bound unspecified and this will become clear later. This is partly motivated by previous work on particle energetics in turbulent flows and is consistent with the phenomenology of Kolmogorov's formulation of structure functions of arbitrary order \cite{kolmogorov_local_1941,sreenivasan_phenomenology_1997}. Using DNS data later in this section, we will verify the validity of this assumption. 

Assuming a power law for each the moments of the relative velocity, we assume that
\begin{equation}
\langle \omega_i\rangle^+ \sim C_i^{1/3}\left(\frac{r}{z}\right)^{1/3},\quad \langle \omega'_i\omega_j'\rangle^+ \sim \varphi_{ij}\left(\mathcal{L}_i\mathcal{L}_j\right)^{1/3}\left(\frac{r}{z}\right)^{2/3},\quad  
\langle {\omega'_i{\omega'_j}^2}\rangle^+ = -\psi_{ij}\left(\mathcal{L}_i\mathcal{L}^2_j\right)^{1/3}\left(\frac{r}{z}\right),
\end{equation}
where $i = \|,z$ and $\varphi_{ij}$ and $\psi_{ij}$ are tensors of the correlation coefficients between longitudinal and vertical relative velocity fluctuations.
Specifically, $\varphi_{ij},\psi_{ij}=1$ if $i=j$ and $-1<\varphi_{ij},\psi_{ij}<1$ otherwise. The coefficients $0<\mathcal{L}_\|<1$, $0<\mathcal{L}_z<1$ are dimensionless parameters that characterize the magnitude of the particle phase energy fluxes (i.e., each component of the particle phase has downscale energy fluxes that are some proportion of the flow phase energy flux at a height $z$). $C_i$ are dimensionless proportionality coefficients characterizing the magnitude of the mean relative velocities. This form assumes a linear relationship between the mean relative velocity and its standard deviation. This scaling behavior has been noted in past studies of particle collisions, where the mean inward relative velocity is a primary quantity of interest \cite{ireland_effect_2016-1}.


We also adopt a power law representation of the radial distribution function
\begin{equation}
\varrho =\mathcal{R}\left(\frac{r}{z}\right)^{-\alpha},  
\end{equation}
where $\mathcal{R}$ and $\alpha$ are constants characterizing the RDF and the degree of spatial clustering in the horizontal plane. It is known that in HIT for non-settling particles, $\varrho$ exhibits a power law behavior in the dissipation range \cite{bragg_new_2014}, though different theories predict slightly different values for the exponents themselves. The dynamics for inertial sub-range clustering take on a more complicated form (see \cite{bragg_mechanisms_2015}) but for the Stokes numbers under consideration in this work, and within a turbulent boundary layer near a solid boundary, the inertial sub-range is not expected to be large enough to render such effects visible. Furthermore, settling is known to modify the power law exponents \cite{bec_gravity-driven_2014,ireland_effect_2016-1}, but in fact it remains that $\varrho$ retains power-law characteristics. 

We must also consider the mixed moments $\langle w'{\omega'_i}^n\rangle$. To estimate these quantities, we make the following assumption based on the triangle inequality:
\begin{equation}
    \langle w'{\omega'_i}^n\rangle \sim \lambda_{i}\left({\langle w'}^2\rangle\right)^{1/2}\left(\langle {\omega_i'}^{2n}\rangle\right)^{1/2},
\end{equation}
where $\lambda_i$ is a proportionality constant.
It follows from \cite{kunkel_study_2006} that the large scale vertical fluctuations are independent of height within the logarithmic layer, giving $\langle {w'}^2\rangle \sim A_0 u_\tau^2$ (recall $A_0$ is a constant). Moreover, we make a Kolmogorov-like assumption that $2n$-th moment of the relative velocity fluctuation is related to the square of the $n$th moment, so it follows that 
\begin{equation}
    \langle w'{\omega'_i}^n\rangle \sim u_\tau^{n+1}\left(\lambda_{i}A_0 \mathcal{L}_i^{n/3}\left(\frac{r}{z}\right)^{n/3}\right).
\end{equation}

Finally, the quantity $\langle w\rangle^+$ is a one point statistical quantity that represents the vertical velocity of particles located at a distance between $r$ and $r+dr$ from a reference particle and is irrespective of the reference particle's velocity. As $r/z$ approaches unity, we expect $\langle w\rangle^+$  to approach the ensemble average vertical settling velocity, which can be larger in magnitude than $\mathrm{Sv}^+$ due to preferential sweeping \cite{wang_settling_1993}. Conversely, as $r/z$ becomes very small, we expect that the turbulent fluctuations will have too short a correlation time to make significant changes to the average particle velocity, so $\langle w\rangle^+\sim \mathrm{Sv}^+$. Notably, if there is no mean drift (i.e. if $\mathrm{Sv}^+\equiv 0$), then $\langle w\rangle^+\equiv 0$. For the current analysis, we assume that $\langle w\rangle^+ = -\beta\mathrm{Sv}^+$, where $\beta$ could be a function or $r/z$ that is bounded below by one and increases to some finite value characterized by the large scales at $r/z \rightarrow 1$. This assumption allows us to characterize the ensemble average settling velocity across all particles (for example, see the work of \cite{grace_reinterpretation_2024}). 
For our argument below, knowledge of the exact magnitudes of these proportionality coefficients is not necessary, as we are only concerned with order of magnitude behavior in $r/z$, but we retain them as they provide information as to what terms may affect the overall tendencies controlling the second order particle structure functions. 

Using these expressions, we can compute each term in \eqref{SoSF} for the longitudinal component (i.e. for $i = \|$)  of the continuum equations. The algebraic details can be found in the Appendix B, and we only report the final scaled expression here: 
\begin{equation}
    \langle {\omega'_\|}^2\rangle^+ \sim \langle \Delta u'_\|\omega'_\|\rangle^+ + \frac{\mathrm{St}^+\mathcal{L}_\|}{z^+}\left(1 - \frac{\alpha}{2} - \frac{2}{3}\left(\frac{C_\|}{\mathcal{L}_\|}\right)^{1/3}+ \mathcal{O}\left(\left(\frac{r}{z}\right)^{2/3}\right)\right). \label{asymptotic longitudinal}
\end{equation}
What we find is that each of $\mathcal{D}_\|$, $\mathcal{M}_\|$, and $\mathcal{F}_\|$ contribute terms at $\mathcal{O}(1)$ and at $\mathcal{O}\left(\left(\frac{r}{z}\right)^{2/3}\right)$. Therefore, in the limit $r/z\ll 1$, we find that the higher moment terms tend towards a constant, which is the minimum longitudinal variance attained by the particle field at equilibrium for a given height, and is due to non-local effects from larger scales. This minimum is proportional to the term
\begin{equation}
     \frac{\mathrm{St}^+\mathcal{L}_\|}{z^+}\left(1 - \frac{\alpha}{2} - \frac{2}{3}\left(\frac{C_\|}{\mathcal{L}_\|}\right)^{1/3} \right), \quad \frac{r}{z}\ll 1.\label{eqn:long flux}
\end{equation}
The minimum has to do with particle inertia, the distance from the wall, and the downscale energy flux (through $\mathcal{L}_\|$), but is modulated by particle clustering through $\alpha$, as well as the mean relative velocity through $C_\|$, both of which decrease the minimum. This term can be interpreted as a mean downscale diffusion driven by clustering \cite{bragg_mechanisms_2015}, and a mean drift independent of the degree of clustering. The form of \eqref{eqn:long flux} implies that the downscale energy flux is persistent but modified by particle clustering. 

Notably, particle settling does not explicitly appear at $\mathcal{O}(1)$, so as we move to smaller and smaller scales, our analysis suggests that the role of settling becomes unimportant, giving further credence that settling is only affected by the large scales of the flow motion. However, the situation is more complicated as settling affects both the mean relative velocity, $\Delta u_i$, and particle clustering, so while there is no explicit $\mathcal{O}(1)$ effect, there is an implicit effect on the downscale energy flux of the particle phase.

The above approximation scheme demonstrates that $\langle {\omega'_\|}^2\rangle$ is controlled in part by $\langle \Delta u'_\|\omega'_\|\rangle^+$ which represents covariance between flow velocity increments and particle velocity increments. This term represents a direct coupling to the flow, and we will not approximate this term here. As we show later, this term is dominant in the $r/z\geq 1$ scaling range. 

The vertical component takes on a similar form, though due to the underlying anisotropy of the flow, the magnitudes of the tendencies will be different (see appendix B for details). The vertical component appears in a similar form to the longitudinal component: 
\begin{multline}
    \langle {\omega'_z}^2\rangle^+ \sim \langle \Delta u'_z\omega'_z\rangle^+ + \frac{\mathrm{St}^+\mathcal{L}_\|}{z^+}\left(\frac{\mathcal{L}_z}{\mathcal{L}_\|}\right)^{2/3}\ldots \\ 
    \left(\left(1 - \frac{\alpha}{2}\right)\psi - \frac{\varphi}{3}\left(\frac{C_z}{\mathcal{L}_z}\right)^{1/3} - \frac{1}{3}\left(\frac{C_\|}{\mathcal{L}_\|}\right)^{1/3}+ \mathcal{O}\left(\left(\frac{r}{z}\right)^{2/3}\right)\right), \label{asymptotic vertical}
\end{multline}
so that in the limit of $r/z\ll 1$, the second term becomes 
\begin{equation}
    \frac{\mathrm{St}^+\mathcal{L}_\|}{z^+}\left(\frac{\mathcal{L}_z}{\mathcal{L}_\|}\right)^{2/3}
    \left(\left(1 - \frac{\alpha}{2}\right)\psi - \frac{\varphi}{3}\left(\frac{C_z}{\mathcal{L}_z}\right)^{1/3} - \frac{1}{3}\left(\frac{C_\|}{\mathcal{L}_\|}\right)^{1/3}\right), \frac{r}{z}\ll 1,
\end{equation}
again highlighting the role of the mean relative velocity (through $C_\|$ and $C_z$) and clustering (through $\alpha$). Importantly, the constant above is different than the longitudinal component necessarily, showing that the vertical relative velocity is necessarily different than the longitudinal as $r/z \rightarrow 0$. 

\begin{figure}
    \centering
    \includegraphics[width=\textwidth]{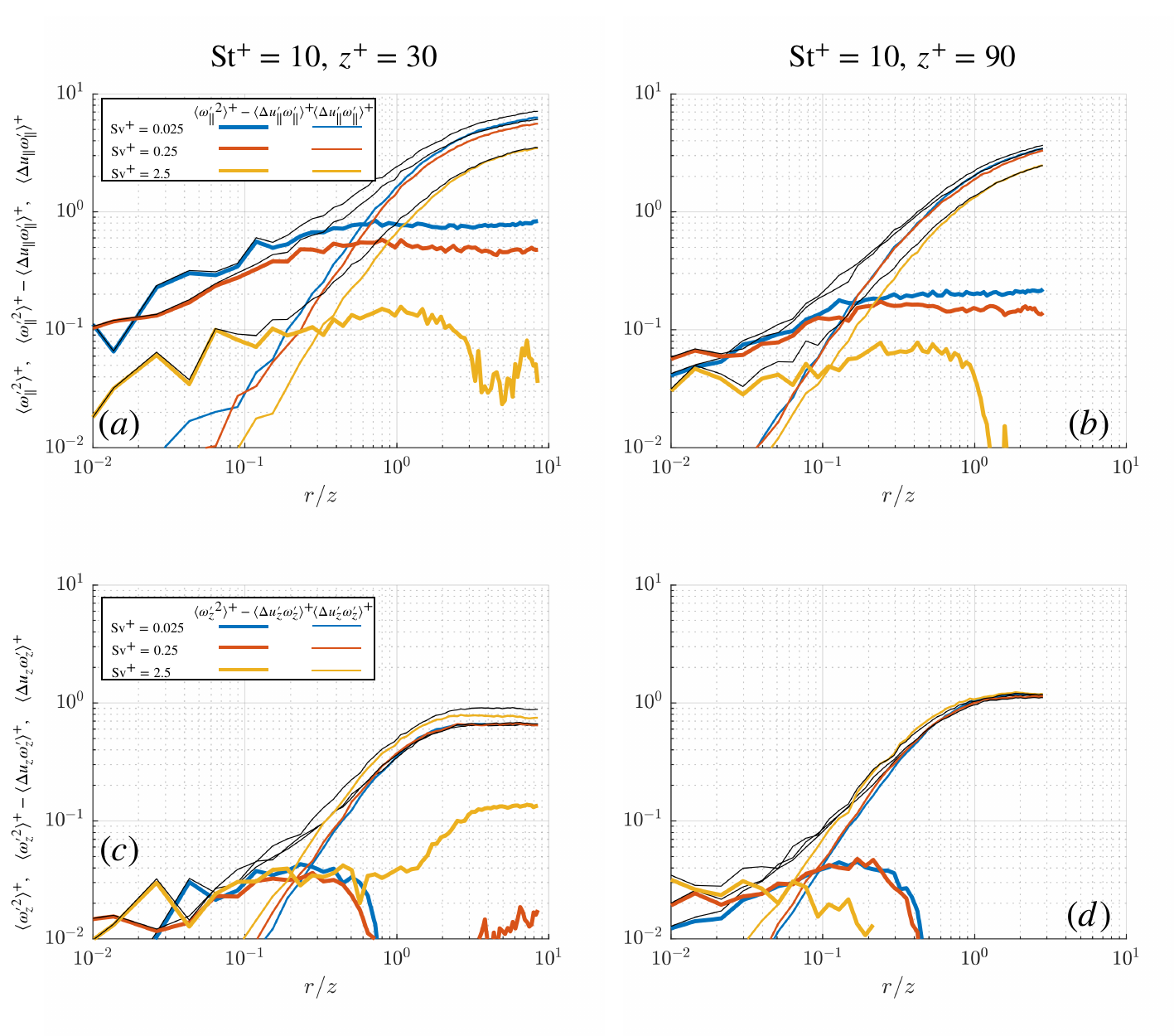}
    \caption{The numerically computed second order structure functions (black curves) for the horizontal (a,b) and vertical (c,d) at $z^+ = 30$ (a,c) and $z^+=90$ (b,d). Note the horizontal axis is extended relative to previous figures to highlight the structure across a wide range of $r/z$. Thin colored curves show $\langle u'_i\omega'_i\rangle^+$, whereas the thick colored curves show $\langle {\omega'_i}^2\rangle^+ - \langle u'_i\omega'_i\rangle^+$, the exact values of the flux terms.}
    \label{fig:Fluxterms}
\end{figure}

We can use the insight gained from the above asymptotic analysis to interpret our DNS results, which are presented below in figure \ref{fig:Fluxterms}. These figures demonstrate the dominant tendencies controlling the structure functions across a wide range of $r/z$ available from the DNS. Figures \ref{fig:Fluxterms}(a,c) show the horizontal (a), and vertical (c), structure functions themselves (depicted by thin black curves), the correlation terms (depicted by thin colored curves), and the flux terms (depicted by thick colored curves) at $z^+ = 30$, while figures \ref{fig:Fluxterms}(b,d) show the same for $z^+=90$. To produce these plots, we directly computed the covariance terms, but as the higher order structure functions require a significant amount of data for statistical convergence, taking numerical derivatives directly would yield a noisy result. To overcome this, we compute the flux terms by computing $\langle {\omega'_i}^2\rangle^+ - \langle \Delta u'_i\omega'_i\rangle^+$. This is suitable for our purposes as no approximations are made in order to derive the model in equation \eqref{SoSF}.

Our DNS results suggest a competition between the covariance terms, $\langle \Delta u'_i\omega'_i\rangle^+$, at large scales, and the flux terms, $\frac{1}{2}\mathrm{St}^+\left(\mathcal{D}_i  + \mathcal{M}_i + \mathcal{F}_i\right)$, at small scales, which is then modified implicitly by particle settling. 
At the large scales the covariance tendencies dominate the overall balance, and tend to fall off rapidly as $r/z\rightarrow 0$. However, the explicit modification by settling to the covariance terms (by increasing $\mathrm{Sv}^+$) is almost negligible.  
While the covariance terms drop out, the flux terms tend to become the leading order tendency controlling the magnitude of the horizontal and vertical structure functions. 
At $r/z\leq 1$, the horizontal component of the flux terms at both heights (figures \ref{fig:Fluxterms}(a,b)) are more strongly modified by the settling, while the vertical component (figures \ref{fig:Fluxterms}(c,d)) shows almost no changes in this same limit. 

The insight here is that the flux terms (quantifying the small scale energy of the particle phase) are noticeably different at small scales, and the difference is decreased as the impact of settling increases (i.e. as  $\mathrm{Sv}^+$ increases).
While settling decreases the flux terms leaving the correlation terms mostly unaffected (or at least affected to a lesser degree), the net effect is that settling changes the dynamic balance controlling the energy exchange in the particle phase. With a larger settling parameter, the particle dynamics are more closely linked to the flow phase dynamics for a larger range of $r/z$. This effect is more evident closer to the wall (figures \ref{fig:Fluxterms}(a,c)) as the magnitude of the second order structure function at small scales tends to decrease with distance from the boundary anyways. This observation is consistent with our asymptotic analysis showing the pre-factor proportional to $(z^+)^{-1}$.

Contrary to the asymptotic analysis, the DNS shows that the flux terms in the horizontal components at both heights have a shallow finite slope as $r/z\rightarrow 0$. While the behavior predicted by the asymptotic analysis suggests a constant behavior, there may still some finite coupling between the particle phase and flow dissipative range dynamics (which our scaling analysis ignores). This changes the scaling behavior as the particles under consideration do not possess a Stokes number large enough to be completely insensitive to dissipation scale flow phase eddies. Moreover, since we have a limited inertial range (if one exists at all), we may not be able to ignore the $\mathcal{O}\left(\left(r/z\right)^{2/3}\right)$ contribution suggested by our analysis.
However, from a qualitative standpoint, it is reassuring that the variation in the flux terms is slow, suggesting that while the scaling of the third moments may change, the qualitative conclusions remain unchanged. 

\begin{figure}
    \centering
    \includegraphics[width=\textwidth]{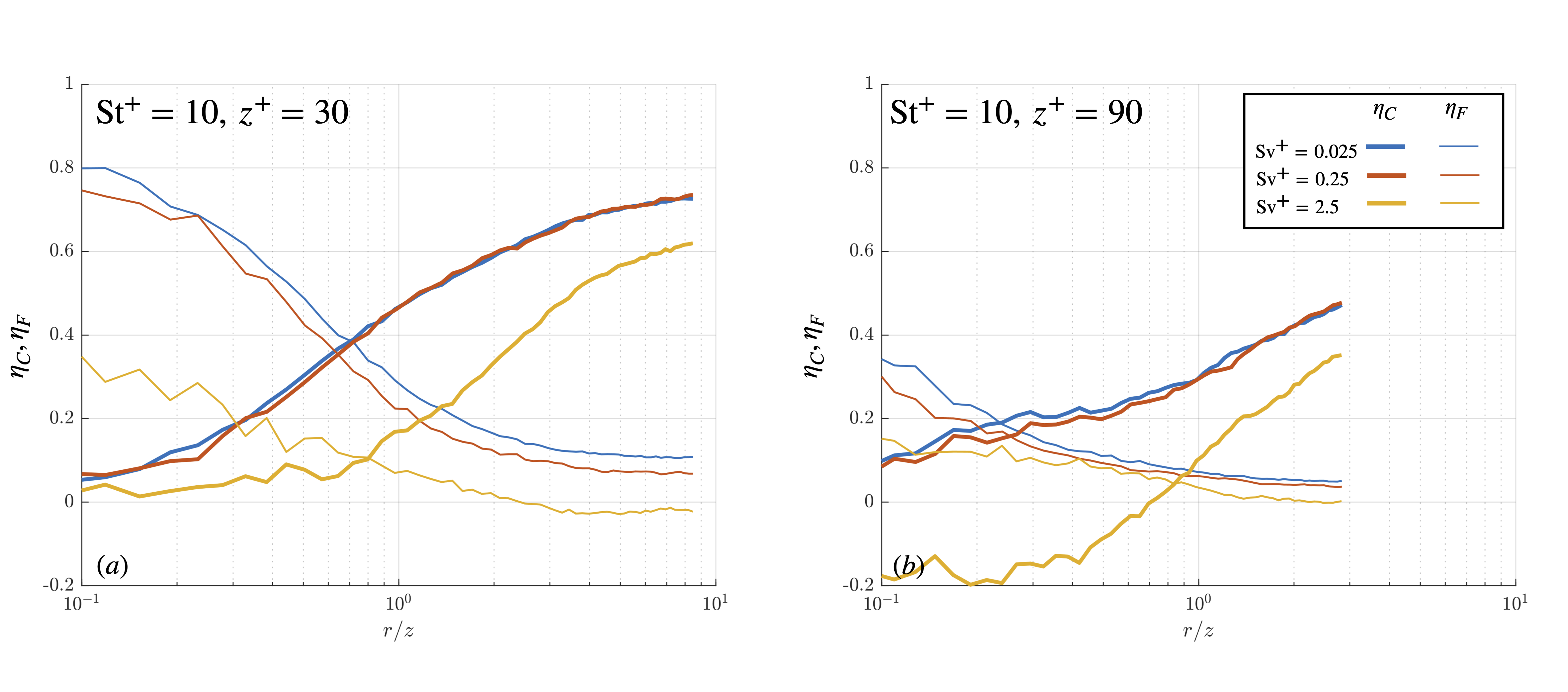}
    \caption{Components of the anisotropy metric, $\eta_C$ and $\eta_F$ (defined in the text) for different values of $\mathrm{Sv}^+$.}
    \label{fig:anisocomponents}
\end{figure}

Using the insight of our analysis above, we finally consider the particle phase anisotropy. 
To analyze the results from our direct numerical simulations, the above asymptotic form motivates us to decompose the anisotropy metric into two components. The first is the difference between the correlation terms, defined as
\begin{equation}
    \eta_C = \frac{\langle \Delta u'_\|\omega'_\|\rangle^+ - \langle \Delta u'_z\omega'_z\rangle^+}{\langle {\omega'_\|}^2\rangle^+ + \langle {\omega'_z}^2\rangle^+}.
\end{equation}
This term describes how the anisotropy changes as the particle phase becomes decoupled from the flow phase as $r/z\rightarrow 0$. The second component, defined as
\begin{equation}
    \eta_F = -\mathrm{St}^+\frac{\left(\mathcal{D}_\| + \mathcal{M}_\| + \mathcal{F}_\|\right) - \left(\mathcal{D}_z + \mathcal{M}_z + \mathcal{F}_z\right)}{\varrho\langle {\omega'_\|}^2\rangle^+ + \varrho\langle {\omega'_z}^2\rangle^+},
\end{equation}
describes the effect of settling and the mean downscale drift on the particle phase anisotropy. Note that $\eta^{(p)} = \eta_C + \eta_F$ exactly. These components are plotted for $z^+ = 30$ in figure \ref{fig:anisocomponents}(a) and $z^+=90$ in figure \ref{fig:anisocomponents}(b).

Figure \ref{fig:anisocomponents} shows the correlation terms (thick curves) and the flux terms (thin curves) for $z^+ = 30$ and $z^+=90$ at fixed $\mathrm{St}^+$ and varied $\mathrm{Sv}^+$. Figure \ref{fig:anisocomponents} shows that the anisotropy at large scales (i.e. $r/z>1$) at both heights is due primarily to the correlation between flow increments and particle velocity increments through $\eta_C$. This verifies that the large scale anisotropy in the particle phase is strongly controlled by the anisotropy of the underlying flow phase. Moreover, we can see that there is very little change to $\eta_C$ by progressively increasing $\mathrm{Sv}^+$, until $\mathrm{Sv}^+ = 2.5$, where it is significantly affected at both heights. Interestingly, at $z^+=90$, $\eta_C$ becomes negative when $r/z<1$ suggesting that settling leads to a faster decorrelation from horizontal components than the vertical. 

When one considers the small scales, the majority of the contribution to the particle phase anisotropy comes from the flux terms. These curves underscore the importance of the non-linear energy cascade (i.e. non-local effects) within the particle phase, and how then almost wholly govern the scale anisotropy. Moreover, since these flux terms are scaled by large scale quantities ($\mathcal{L}_i$), the implication is that the small scale anisotropy is influenced by the large scale response of the particle phase to the fluid phase. Figure \ref{fig:anisocomponents}(a) shows significant differences in $\eta_F$ at $z^+ = 30$, suggesting that the cascade of energy in the particle phase is significantly modified by settling at high $\mathrm{Sv}^+$, thus rendering the small scales only weakly anisotropic. Interestingly, in figure \ref{fig:anisocomponents}(b), the effect of settling on the small scale anisotropy is smaller. This is evident in equation \eqref{asymptotic longitudinal}, where $\eta_F$ is proportional to $(z^+)^{-1}$, meaning a smaller effect further from the boundary. Overall, the small scale anisotropy is only weakly modified by particle settling when one considers distances far enough from the wall. 


\section{Discussion and Conclusions}
\label{sect:discussion}

In this work, we considered the broad question of how anisotropy present in a wall bounded turbulent flow is imparted onto a field of Lagrangian particles. We defined both an Eulerian scalewise anisotropy for both the fluid and the dispersed phases by employing second order structure functions (SoSF) in $r-z$ space. This technique allowed us to make a direct comparison of the energetic anisotropy of both phases, and how their characteristics differ with scale. We focused our analysis on two regions of the flow: the logarithmic layer (the self-similar region of the flow), and the buffer layer, where the underlying flow phase anisotropy is strong across all scales of motion. To analyze our system, we derived a theoretical model for the tendencies controlling the particle phase second order structure functions, and despite the relatively low Reynolds number, we found qualitative agreement between an asymptotic scaling argument for the the particle phase SoSF and Eulerian-Lagrangian coupled DNS in this region, suggesting that our results may scale to higher Reynolds numbers.

The main conclusion from this work is that for sufficiently inertial Lagrangian point particles, the particle phase demonstrates a significant level of anisotropy in the formally isotropic dissipative range of the flow phase. We hypothesize that this is driven by the multi-scale inertial response of particles to the large scale anisotropy of the flow. These so-called non-local effects (a downscale drift) lead to increased energy in the at small scales relative to purely passive particles and has been noted in previous studies. The novelty here is that since the large scales of the flow are strongly anisotropic (an effect increasing with Reynolds number), the horizontal and vertical energy densities of the dispersed phase at small scales were anisotropic themselves. By computing the scalewise anisotropy, we found that the small scale (i.e. $r/z\ll 1$) anisotropy was comparable to the large scale anisotropy (i.e. $r/z >1$), with a minimum in between. 

Our asymptotic analysis found that the role of particle settling in the small scales was minor until small separations (i.e. $r/z\ll 1$), and was unsurprisingly strongest when $\mathrm{Sv}^+$ became order unity for $\mathrm{St}^+=10$. 
We found that as $\mathrm{Sv}^+$ was increased, this preserved anisotropy at the small scales began to decrease, and we showed that this is due to particle settling directly decreasing the minimum horizontal variance at small separations more than the vertical. By considering $\mathrm{Sv}^+=2.5$, we see that the anisotropy of the small scales actually becomes negative, suggesting a fundamental change in the topology of dispersed phase ``eddies". The implication is that that since the small scales in the particle phase are already positively anisotropic (i.e. fluctuations have a pancake-like topology), settling works to draw the small scales to isotropy, and continually increasing $\mathrm{Sv}^+$ eventually leads to a slightly negative anisotropy (fluctuations have a column-like topology).  

This work also brings about several interesting considerations for a potential prognostic Eulerian model for inertial settling particles. Our results point to a more general phenomenon that while the fluid phase may return to isotropy at small enough scales (i.e., in the inertial subrange and dissipative ranges), the emergence of non-local effects due to particle inertia may imply more generally that the large scale anisotropy of the flow state is fossilized at smaller scales in the particle field. This multi-scale effect could have interesting and important consequences for properly modeling both one- and two-way coupled phenomena across a wide range of applications. For example, particle collision rates (which inherently depend upon the small scale statistics of the particle field) may require information regarding the large scale flow state.

Another consideration may be derived from the practical concerns associated with the choice of numerical method for modeling the particle phase as a continuum. For example, a diffusivity may be required for numerical solution stability (as is often a necessity for high order spectral methods), or may be implicit in the method itself, such as for lower order finite difference or finite volume methods. Therefore, to reproduce the small scale statistics of a multi-phase continuum model (i.e. the particle phase energy cascade), care must be taken to pick a particle diffusivity small enough such that the minimum energy is not too strongly attenuated while still preserving model stability. One potential approach may be to use a LES-style closure for the particle phase, making sure to appropriately model the solid phase energy flux from the resolved scales. This is especially relevant for appropriately modeling the subgrid scale dynamics of Eulerian-Lagrangian LES, which are a popular middle ground between small scale DNS and operational scale models (\cite{park_simple_2017} and references therein). 

A final consequence is how energy is passed from the particle phase back to the flow phase in a two-way coupled sense. There is significant evidence to suggest that the two way coupling between solid particles and the flow phase may either augment or attenuate turbulence introduce energy at the small scales and attenuate energy at larger scales (known as pivoting \cite{hassaini_scale--scale_2022, poelma_particle-turbulence_2006}. Our work suggests a potential hypothesis for this, whereas since the energy of the particle phase is much higher than the flow phase in the dissipative range, the back reaction onto the flow phase by the particles may scale with the minimum energy attained by the particle phase due to non-local effects (i.e. the flux terms described above), thereby implying that energy is injected directly into dissipative range. 

The findings of this study also highlight that the energy exchange between the solid phase and the fluid phase in this range may not be isotropic, since the energy flux in each component is different (due to the large scale anisotropy). Since our results highlight that the small scale energetic content of the horizontal component is higher on average than in the vertical component, two-way coupling in anisotropic turbulence may in fact place a much larger drag onto the horizontal component, thus warping their characteristic shapes more than the vertical component. For example, \cite{hassaini_scale--scale_2022} comment that settling particles at sufficient loading tend to warp eddies in the vertical as potential energy from the particle phase is passed to kinetic energy, and then onto the flow, thereby weakening horizontal fluctuations and adding energy to vertical ones. This could lead to fundamentally different energy exchange pathways in anisotropic turbulence, versus isotropic turbulence, but a dedicated study is required to test this hypothesis.

\begin{acknowledgments}

The authors would like to acknowledge Grant No. W911NF2220222 from the U.S. Army Research Office, and the Center for Research Computing at the University of Notre Dame and the University of Colorado Research Computing Office. AG would also like to acknowledge partial funding by the Natural Science and Engineering Research Council of Canada number PDF - 587364 - 2024. 
\end{acknowledgments}

\appendix

\section{Derivation of second order structure functions}

This appendix uses techniques from kinetic theory to derive the continuum equations for the second order relative velocity moments of the particle phase JPDF. These second order moment corresponds to the second order structure functions of the particle phase velocity. 

We first defined the JFPF as 
\begin{equation}
    \mathcal{P} = \mathcal{P}(r,z,w,\omega_\|,\omega_z),
\end{equation}
which represents the probability density of a pair of particles existing in a hypercube in phase space. Recall that $r$ represents the longitudinal separation between two particles, $z$ represents the height of a reference particle, $w$ is the vertical veloctiy of a reference particle, $\omega_\|$ is the longitudinal relative velocity between satellite and reference particles, and $\omega_z$ is the relative vertical velocity of satellite and reference particles. Assuming that we can neglect particle collisions, the JPDF is a conserved quantity in phase space and we can take the time derivative of $\mathcal{P}$ to write down the master transport equation
\begin{equation}
    \frac{\partial \mathcal{P}}{\partial t} +\frac{\partial}{\partial z}w\mathcal{P} + \frac{1}{r}\frac{\partial}{\partial r}r\omega_\|\mathcal{P} + \frac{\partial }{\partial \omega_\|}\dot{\omega}_\|\mathcal{P}+ \frac{\partial }{\partial \omega_z}\dot{\omega}_z\mathcal{P} = 0.
\end{equation}
For the following, we ignore time dependence in all quantities as we are concerned only with the steady state dynamics. Note that $\dot{\omega}_\|$ and $\dot{\omega}_z$ are defined by their configuration space equations in the main text.

A useful intermediate step is to derive the equation for the zeroth moment by integrating the above equations across all longitudinal and vertical relative velocities. This equation will be used to simplify the equation of the second order moments, and represents a conservation equation for the zeroth relative velocity moment of the JPDF:
\begin{equation}
    \frac{\partial}{\partial z}\langle w\rangle\varrho + \frac{1}{r}\frac{\partial }{\partial r}r\langle \omega_\|\rangle\varrho = 0 \label{zeroth}
\end{equation}
where the marginal JDF is defined by
\begin{align}
    \varrho = \int_{\mathbb{R}^3}\mathcal{P}d\omega_\|\omega_z dw.
\end{align}
The marginal is related to the probability of two particles separated by a longitudinal distance $r$ where one is located at a height $z$ (i.e. the longitudinal RDF centered at a height $z$).

The averages of the vertical particle velocity and the longitudinal separation velocity conditioned on the remaining phase space variables $r$ and $z$ (denoted by $\langle\cdot\rangle$) are defined as 
\begin{equation}
    \varrho \langle w\rangle = \int_{\mathbb{R}^3}w\mathcal{P}d\omega_\|d\omega_zdw,\quad \varrho \langle \omega_\|\rangle = \int_{\mathbb{R}^3}\omega_\|\mathcal{P}d\omega_\|d\omega_z dw.
\end{equation}
The first term represents the average vertical velocity of particles such that they are a longitudinal distance away from a reference particle irrespective of the reference particle's velocity (i.e. $\langle w\rangle$ is a one-point quantity). The second quantity is the mean longitudinal relative velocity of all particles.

Equation \eqref{zeroth} indicates that if $\varrho\langle w\rangle$ is dependent upon height (say due to particle settling, and dynamic effects such as preferential sweeping and turbophoresis \cite{grace_reinterpretation_2024}), then there is a non-zero mean longitudinal relative velocity in general. Therefore, we must account for both the mean and the fluctuating components of the longitudinal relative velocities and satellite (or reference) particle velocities.

First, we make s variable substitutions $\omega_i = \omega'_i + \langle \omega_i\rangle$ and $w = w' + \langle w\rangle$, so that $\omega'_i$ and $w'$ are new phase space variables while $\langle \omega_i\rangle$ and $\langle w\rangle$ represent the conditional average on $r$ and $z$ (and are thus dependent only upon those variables). Under this change of variables, the derivatives with respect to $\omega_i$ become
\begin{equation}
   \frac{\partial}{\partial \omega_\|} \rightarrow \frac{\partial}{\partial \omega'_\|},
\end{equation}
whereas the $r$ and $z$ derivatives become (by the chain rule from calculus)
\begin{equation}
   \frac{\partial}{\partial z} \rightarrow \frac{\partial}{\partial z} - \frac{\partial \langle\omega_i\rangle}{\partial z}\frac{\partial}{\partial \omega'_i} ,\quad \frac{1}{r}\frac{\partial}{\partial r}r \rightarrow \frac{1}{r}\frac{\partial}{\partial r}r - \frac{\partial \langle\omega_i\rangle}{\partial r}\frac{\partial}{\partial \omega'_i},
\end{equation}
where the repeated subscript $i$ implies summation over the longitudinal and vertical components.
Upon making the above substitution, the master equation becomes
\begin{multline}
        \frac{\partial}{\partial z}w'\mathcal{P} + \frac{1}{r}\frac{\partial}{\partial r}r\omega'_\|\mathcal{P} + \frac{\partial }{\partial \omega'_\|}\dot{\omega}'_\|\mathcal{P}+ \frac{\partial }{\partial \omega'_z}\dot{\omega}'_z\mathcal{P} = \ldots \\ 
        + \frac{\partial \langle\omega_i\rangle}{\partial z}\frac{\partial}{\partial \omega'_i}w\mathcal{P} + \frac{\partial \langle\omega_i\rangle}{\partial r}\frac{\partial}{\partial \omega'_i}\omega_\|\mathcal{P} -\frac{\partial}{\partial z}\langle w\rangle\mathcal{P} - \frac{1}{r}\frac{\partial }{\partial r}r\langle\omega_\|\rangle\mathcal{P} \ldots \\ 
        -\langle \dot{\omega}_\|\rangle\frac{\partial}{\partial \omega'_\|}\mathcal{P} - \langle \dot{\omega}_z\rangle\frac{\partial}{\partial \omega'_z}\mathcal{P},
\end{multline}

where all terms on the left hand side on hold have to do with fluctuating variables, while terms on the right hand side each have some influence from mean quantities. Note that mean and fluctuating components of $\dot{\omega_i}$ are defined as 
\begin{align}
    \langle \dot{\omega}_i\rangle = \frac{\langle\Delta u_i\rangle_{w,\omega_i} - \langle \omega_i\rangle}{\tau_p} \\ 
    \dot{\omega}'_i = \frac{\langle\Delta u_i'\rangle_{w,\omega_i} - \omega'_i}{\tau_p},
\end{align}
where the notation $\langle \cdot \rangle_{w,\omega_i}$ represents an average sampled relative velocity conditioned on all phase space variables. See \cite{reeks_development_2021} for an explanation.

To compute the equations for the second order particle phase structure function, we multiply the master equation by ${\omega'_i}^2$ and integrate over all relative velocity fluctuations and satellite particle velocity fluctuations. Noting that any terms multiplied by a first order fluctuating quantity integrate to zero, the master equation simplifies quite significantly to the following form
\begin{equation}
        \varrho\langle {\omega'_\|}^2\rangle = \varrho\langle \Delta u'_\|\omega'_\|\rangle - \frac{\tau_p}{2}\left(\mathcal{D}_\| + \mathcal{M}_\| + \mathcal{F}_\|\right).
\end{equation}
In the above, $\mathcal{D}_\|$ can be thought of as a mean transport term and is defined as 
\begin{equation}
    \mathcal{D}_\| = \varrho\langle w\rangle \frac{\partial}{\partial z}\langle {\omega'_\|}^2\rangle + \varrho\langle \omega_\|\rangle \frac{\partial}{\partial r}\langle {\omega'_\|}^2\rangle,
    \end{equation}
$\mathcal{F}_\|$ is defined as
\begin{equation}
    \mathcal{F}_\| = \frac{\partial}{\partial z}\varrho \langle w'{\omega'_\|}^2\rangle + \frac{1}{r}\frac{\partial}{\partial r}r\varrho \langle {\omega'_\|}^3\rangle,
\end{equation}
and $\mathcal{M}_\|$ represents an additional contribution from the mean
\begin{equation}
   \mathcal{M}_\| = 2\varrho\langle w' \omega'_\|\rangle \frac{\partial}{\partial z}\langle \omega_\|\rangle + 2\varrho\langle {\omega'}^2_\|\rangle \frac{\partial}{\partial r}\langle \omega_\|\rangle
\end{equation}

Likewise, following an identical series of manipulations, the vertical component is 
\begin{equation}
        \varrho\langle {\omega'_z}^2\rangle = \varrho\langle \Delta u'_z\omega'_z\rangle - \frac{\tau_p}{2}\left(\mathcal{D}_z + \mathcal{M}_z + \mathcal{F}_z\right).
\end{equation}
The vertical transport term is 
\begin{equation}
    \mathcal{D}_z = \varrho\langle w\rangle \frac{\partial}{\partial z}\langle {\omega'_z}^2\rangle + \varrho\langle \omega_\|\rangle \frac{\partial}{\partial r}\langle {\omega'_z}^2\rangle,
    \end{equation}
$\mathcal{F}_z$ is defined as
\begin{equation}
    \mathcal{F}_z = \frac{\partial}{\partial z}\varrho \langle w'{\omega'_z}^2\rangle + \frac{1}{r}\frac{\partial}{\partial r}r\varrho \langle {\omega'_\|}{\omega'_z}^2\rangle,
\end{equation}
and 
$\mathcal{M}_z$ is defined as
\begin{equation}
     \mathcal{M}_z = 2\varrho\langle w' \omega'_z\rangle \frac{\partial}{\partial z}\langle \omega_z\rangle + 2\varrho\langle \omega'_\|{\omega'}_z\rangle \frac{\partial}{\partial r}\langle \omega_z\rangle  
\end{equation}

\section{Asymptotic approximation for flux terms}

This appendix derives the flux terms given the power law assumptions described in the main manuscript and are assumed to hold in the limit $r/z \ll 1$ and $r/z \gg $ viscous should it exist. Finally, we are assuming the above relationships hold in the asymptotic sense (i.e. $r/z \ll 1$) which is why we use  `$\sim$'

\subsection{Power law assumptions of mean quantities}
We make the following assumptions for the mean two point quantities:
\begin{equation}
\langle \omega_i\rangle \sim u_\tau C_i^{1/3}\left(\frac{r}{z}\right)^{1/3},\quad \langle \omega'_i\omega_j'\rangle \sim u_\tau^2\varphi_{ij}\left(\mathcal{L}_i\mathcal{L}_j\right)^{1/3}\left(\frac{r}{z}\right)^{2/3},\quad  
\langle {\omega'_i{\omega'_j}^2}\rangle = -u_\tau^3\psi_{ij}\left(\mathcal{L}_i\mathcal{L}^2_j\right)^{1/3}\left(\frac{r}{z}\right),
\end{equation}
where $\varphi_{ij}$ and $\psi_{ij}$ are tensors of the correlation coefficients between longitudinal and vertical relative velocity fluctuations.
Specifically, $\varphi_{ij},\psi_{ij}=1$ if $i=j$ and $-1<\varphi_{ij},\psi_{ij}<1$ otherwise. The coefficients $\mathcal{L}_i$ and $C_i$ are defined in the main text. Recall that we also adopt a power law representation of the radial distribution function
\begin{equation}
\varrho =\mathcal{R}\left(\frac{r}{z}\right)^{-\alpha},  
\end{equation}
with $\mathcal{R}$ and $\alpha$ discussed in the main text.

We must also consider the mixed moments $\langle w'{\omega'_i}^n\rangle$. To estimate these quantities, we make the following assumption based on the triangle inequality;
\begin{equation}
    \langle w'{\omega'_i}^n\rangle \sim \lambda_{i}\left({\langle w'}^2\rangle\right)^{1/2}\left(\langle {\omega_i'}^{2n}\rangle\right)^{1/2},
\end{equation}
and it follows from \cite{kunkel_study_2006} that since the large scale vertical fluctuations are independent of height within the logarithmic regime, we have $\langle {w'}^2\rangle \sim A_0 u_\tau^2$ (recall $A_0$ is a constant), and it follows that 
\begin{equation}
    \langle w'{\omega'_i}^n\rangle \sim u_\tau^{n+1}\left(\lambda_{i}A_0 \mathcal{L}_i^{n/3}\left(\frac{r}{z}\right)^{n/3}\right).
\end{equation}
Finally, as discussed in the text, we make the assumption that $\langle w\rangle \sim -\beta \mathrm{Sv}^+$, where $\beta$ may be a function of function of $r/z$ with a positive exponent, but for we leave it undetermined for the sake of the following analysis.

\subsection{Asymptotic analysis}
First, we consider the terms $\mathcal{M}_\|$ and $\mathcal{M}_z$, defined as
\begin{align}
    \mathcal{M}_\| &= 2\varrho\langle w'\omega'_\|\rangle\frac{\partial}{\partial z}\langle\omega_\|\rangle + 2\varrho\langle {\omega'_\|}^2\rangle\frac{\partial}{\partial r}\langle\omega_\|\rangle \\
    \mathcal{M}_z &= 2\varrho\langle w'\omega'_z\rangle\frac{\partial}{\partial z}\langle\omega_z\rangle + 2\varrho\langle {\omega'_\|\omega'_z}\rangle\frac{\partial}{\partial r}\langle\omega_z\rangle.
\end{align}
Taking the $z$ and $r$ derivatives of $\langle \omega_i\rangle$, we find that
\begin{equation}
    \frac{\partial}{\partial z}\langle\omega_i\rangle \sim -\frac{1}{3}\frac{u_\tau}{z} C_i^{1/3}\left(\frac{r}{z}\right)^{1/3},
\end{equation}
whereas
\begin{equation}
    \frac{\partial}{\partial r}\langle\omega_i\rangle \sim \frac{1}{3}\frac{u_\tau}{r} C_i^{1/3}\left(\frac{r}{z}\right)^{1/3}.
\end{equation}
Focusing first on $\mathcal{M}_\|$, substituting our expressions for $\langle w'\omega'_\|\rangle$, $\langle {\omega'_\|}^2\rangle$, and $\langle \omega_\|\rangle$ we have
\begin{equation}
    \mathcal{M}_\| \sim -\frac{2}{3}\varrho u_\tau^3\frac{\lambda_\| A_0 \mathcal{L}^{1/3}_\|C_\|^{1/3}}{z} \left(\frac{r}{z}\right)^{2/3} + \frac{2}{3}\varrho u_\tau^3\frac{\mathcal{L}_\|^{2/3}C_\|^{1/3}}{r}\left(\frac{r}{z}\right).
\end{equation}
Simplifying the above by removing a common factor from both terms, we arrive at the following simplified expression for $\mathcal{M}_\|$
\begin{equation}
    \mathcal{M}_\| \sim \frac{2}{3}\varrho u_\tau^3 \frac{\mathcal{L}^{2/3}_\|C_\|^{1/3}}{z}\left( 1 - \gamma_\|\left(\frac{r}{z}\right)^{2/3}\right),
\end{equation}
where we have grouped constants to define 
\begin{equation}
    \gamma_\| = \frac{A_0\lambda_\|}{\mathcal{L}_\|^{1/3}}.
\end{equation}
Likewise, following the same series of manipulations, but instead using $\langle w'\omega'_z\rangle$, $\langle {\omega'_\|\omega'_z}\rangle$, and $\langle \omega_z\rangle$, we arrive at a similar expression for $\mathcal{M}_z$
\begin{equation}
    \mathcal{M}_z \sim  \frac{2}{3}\varrho u_\tau^3 \frac{\left(\mathcal{L}_\|\mathcal{L}_z C_z\right)^{1/3}}{z}\left(\varphi -\gamma_z\left(\frac{r}{z}\right)^{2/3}\right),
\end{equation}
where $\gamma_z$ is defined as 
\begin{equation}
    \gamma_z = \frac{A_0\lambda_{z}}{\mathcal{L}^{1/3}_\|}
\end{equation}
noting that there are differences in magnitude at both $\mathcal{O}(1)$ and $\mathcal{O}\left(\left(r/z\right)^{2/3}\right)$ relative to $\mathcal{M}_\|$.

Next, we consider the transport terms defined as follows (for the $i$th component)
\begin{equation}
    \mathcal{D}_i =  \varrho \langle w\rangle\frac{\partial }{\partial z}\langle {\omega_i'}^2\rangle + \langle \omega_\|\frac{\partial}{\partial r}\langle {\omega_i'}^2\rangle.
\end{equation}
Taking the $z$ and $r$ derivatives of $\langle {\omega'_i}^2\rangle$, we find that
\begin{equation}
    \frac{\partial}{\partial z}\langle {\omega'_i}^2\rangle \sim -\frac{2}{3}\frac{u^2_\tau}{z} \mathcal{L}_i^{2/3}\left(\frac{r}{z}\right)^{2/3},
\end{equation}
whereas
\begin{equation}
    \frac{\partial}{\partial r}\langle {\omega'_i}^2\rangle \sim \frac{2}{3}\frac{u^2_\tau}{r} \mathcal{L}_i^{2/3}\left(\frac{r}{z}\right)^{2/3}.
\end{equation}
Substituting in the definitions for $\langle \omega_\|\rangle$ and $\langle w\rangle$, we arrive at 
\begin{equation}
    \mathcal{D}_i\sim \frac{2}{3}\varrho\frac{u^2_\tau \beta v_g\mathcal{L}_i^{2/3}}{z} \left(\frac{r}{z}\right)^{2/3} +  \frac{2}{3}\varrho\frac{u^3_\tau C_\|^{1/3}\mathcal{L}_i^{2/3}}{z},
\end{equation}
where we have canceled the common $r$ in the numerator and denominator in the second term above.
After some rearrangement, the above is written as
\begin{equation}
    \mathcal{D}_i\sim \frac{2}{3}\varrho u^3_\tau\frac{\mathcal{L}_i^{2/3}C_\|^{1/3}}{z}\left(1 +  \frac{\beta \mathrm{Sv}^+}{C_\|^{1/3}}\left(\frac{r}{z}\right)^{2/3}\right).
\end{equation}

Finally, we consider the flux terms defined as follows,
\begin{align}
    \mathcal{F}_\| &= \frac{\partial}{\partial z}\varrho \langle w'{\omega'_\|}^2\rangle + \frac{1}{ r}\frac{\partial}{\partial r}r\varrho\langle {\omega'_\|}^3\rangle, \\ 
     \mathcal{F}_z &= \frac{\partial}{\partial z}\varrho \langle w'{\omega'_z}^2\rangle + \frac{1}{r}\frac{\partial}{\partial r}r\varrho\langle {\omega'_\|}{\omega'_z}^2\rangle,   
\end{align}
Importantly, $\varrho$ is located within the derivatives, meaning we must evaluate the derivatives of the following two terms
\begin{align}
    \varrho \langle {\omega'_\|{\omega'_j}^2}\rangle &= -u_\tau^3\mathcal{R}\psi_{\|j}\left(\mathcal{L}_\|\mathcal{L}^2_j\right)^{1/3}\left(\frac{r}{z}\right)^{1-\alpha} \\ 
    \varrho \langle {w'{\omega'_j}^2}\rangle &= u_\tau^3\mathcal{R}\lambda_jA_0\mathcal{L}^{2/3}_j\left(\frac{r}{z}\right)^{2/3-\alpha}
\end{align}
Taking the $z$ derivative of $ \varrho \langle {w'{\omega'_j}^2}\rangle$, we find that
\begin{equation}
    \frac{\partial}{\partial z} \varrho \langle {w'{\omega'_j}^2}\rangle \sim -\left(\frac{2}{3} -\alpha\right)\varrho u^3_\tau\frac{\lambda_jA_0\mathcal{L}_j^{2/3}}{z}\left(\frac{r}{z}\right)^{2/3},
\end{equation}
where we have re-substituted in the definition for $\varrho$. Likewise, taking the radial derivative of $\varrho \langle {\omega'_\|}{\omega'_j}^2\rangle$,
\begin{equation}
    \frac{1}{r}\frac{\partial}{\partial r}r\varrho \langle {\omega'_\|}{\omega'_j}^2\rangle \sim -(2 - \alpha)\varrho u^3_\tau\frac{\psi_{\|j}\left(\mathcal{L}_\|\mathcal{L}_j^2\right)^{1/3}}{z}.
\end{equation}

Plugging the all third order terms into the above equations, we get the following estimates for the magnitudes of these terms
\begin{align}
    \mathcal{F}_\| &= -\varrho u_\tau^3\frac{\mathcal{L}_\|}{z}\left((2-\alpha) + \left(\frac{2}{3} - \alpha\right)\gamma_\|\left(\frac{r}{z}\right)^{2/3} \right), \\ 
    \mathcal{F}_z &= -\varrho u_\tau^3\frac{\mathcal{L}_\|}{z}\left(\frac{\mathcal{L}_z}{\mathcal{L}_\|}\right)^{2/3}\left((2-\alpha)\psi + \left(\frac{2}{3} - \alpha\right)\gamma_z\left(\frac{r}{z}\right)^{2/3} \right).   
\end{align}

\bibliography{zLibrary}

\end{document}